\documentclass[superscriptaddress,twocolumn,amsmath,amssymb]{revtex4-1}
\pdfoutput=1
\usepackage{graphicx}
\usepackage{eurosym}
\usepackage{amsmath}
\usepackage{amssymb}
\usepackage{dcolumn}
\usepackage{bm}
\usepackage{subfigure}
\usepackage{soul}
\usepackage{algcompatible}
\usepackage{newfloat}
\usepackage{comment}
\DeclareFloatingEnvironment[
    fileext=loa,
    listname=List of Algorithms,
    name=ALGORITHM,
    placement=tbhp,
]{algorithm}

\def\avg#1{\mathinner{\langle{#1}\rangle}}
\def\bra#1{\mathinner{\langle{#1}|}}
\def\ket#1{\mathinner{|{#1}\rangle}}
\newcommand{\braket}[2]{\langle #1|#2\rangle}

\DeclareMathOperator{\Tr}{Tr}

\newcommand{\ignore}[1]{}

\newcommand{\be}{\begin{equaArXivtion}}
\newcommand{\ee}{\end{equation}}
\newcommand{\ba}{\begin{eqnarray}}
\newcommand{\ea}{\end{eqnarray}}

\begin{document}

\title{Hybrid Quantum-Classical Hierarchy for Mitigation of Decoherence and Determination of Excited States}

\author{Jarrod R. McClean} 
\email[Corresponding author: ]{jmcclean@lbl.gov}
\affiliation{Computational Research Division, Lawrence Berkeley National Laboratory, Berkeley, CA 94720, USA}

\author{Mollie E. Schwartz} 
\affiliation{Quantum Nanoelectronics Laboratory, Department of Physics, University of California, Berkeley, Berkeley, CA 94720, USA}

\author{Jonathan Carter} 
\affiliation{Computational Research Division, Lawrence Berkeley National Laboratory, Berkeley, CA 94720, USA}

\author{Wibe A. de Jong} 
\affiliation{Computational Research Division, Lawrence Berkeley National Laboratory, Berkeley, CA 94720, USA}

\begin{abstract}Using quantum devices supported by classical computational resources is a promising approach to quantum-enabled computation.  One example of such a hybrid quantum-classical approach is the variational quantum eigensolver (VQE) built to utilize quantum resources for the solution of eigenvalue problems and optimizations with minimal coherence time requirements by leveraging classical computational resources.  These algorithms have been placed among the candidates for first to achieve supremacy over classical computation.  Here, we provide evidence for the conjecture that variational approaches can automatically suppress even non-systematic decoherence errors by introducing an exactly solvable channel model of variational state preparation.  Moreover, we show how variational quantum-classical approaches fit in a more general hierarchy of measurement and classical computation that allows one to obtain increasingly accurate solutions with additional classical resources.  We demonstrate numerically on a sample electronic system that this method both allows for the accurate determination of excited electronic states as well as reduces the impact of decoherence, without using any additional quantum coherence time or formal error correction codes.
\end{abstract}

\maketitle


First conceived of by Richard Feynman~\cite{Feynman1982}, quantum computers have the potential to offer radical advances in solving important problems ranging from optimization and eigenvalue problems to materials design.  One problem of particular recent interest is that of quantum chemistry, where quantum computers have the potential to offer an exponential speedup in the determination of physical and chemical properties~\cite{Aspuru:2005,Kassal:2011,Huh:2015}.  This problem has received attention both because of its great practical utility, and because it is believed that it may be one of the first approaches to demonstrate the superiority of a quantum computer over currently available classical computers~\cite{Wecker:2015a,Mueck:2015}.

Recently, there have been a number of advances in quantum chemistry on quantum computers both algorithmically and technologically.  The original work utilized the quantum phase estimation algorithm~\cite{Abrams1997,Abrams1999,Kitaev:1997} and analyzed the use of adiabatic state preparation in chemical problems.  Since then, the cost of the quantum phase estimation procedure has been brought down dramatically through considerations of physical locality of interactions, chemical insights, and more general algorithmic enhancements~\cite{McClean:2014,Babbush:2015,Hastings2014,Poulin2014,Babbush:2015a}.  Additionally, prototype implementations of many of these algorithms have now been verified in the lab on quantum technologies such as quantum photonics, ion traps, NMR computers, and nitrogen vacancies in diamond ~\cite{Lanyon:2010,Lu:2011,Walther:2012,Peruzzo2014,Wang:2015,Shen:2015}.
\begin{figure}[t!]
\includegraphics[width=8.0 cm]{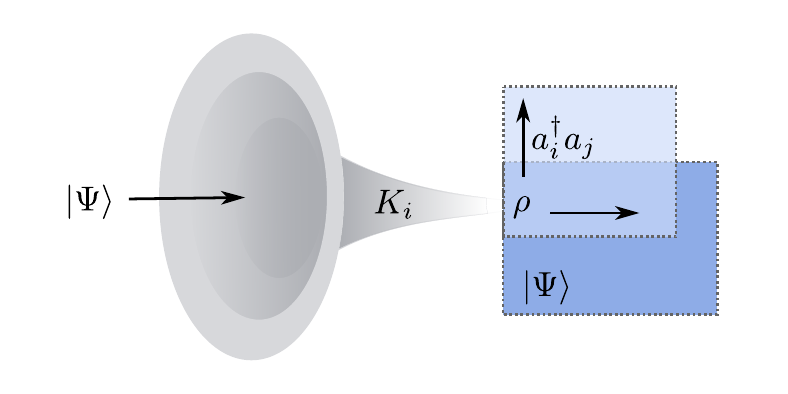}
\caption{A cartoon schematic of the quantum subspace expansion proposed in this work.  One prepares a quantum state $\ket{\Psi}$ that is passed through a quantum channel.  At the exit of the channel, partial tomography of the state is used to expand in a linear subspace around the resulting quantum state.  This subspace is used to determine both the ground and excited states of a quantum Hamiltonian of interest while  also potentially correcting for errors caused by the quantum channel. \label{fig:QSESummarySchematic}}
\end{figure}

While there have been significant developments in quantum hardware across a variety of platforms, many of these algorithms cannot be faithfully run on current or near-future technology.  To combat this problem, a hybrid quantum classical approach was developed, with the the idea that quantum processors should only be used for tasks that have a strong comparative advantage in the quantum domain.~\cite{Peruzzo2014}. Hybrid quantum-classical variational approaches work analogously to classical variational approaches, by preparing a parameterized ansatz on the quantum device and minimizing the energy with respect to the parameters.  The use of a quantum device expands the classes of ansatz one may explore, including many which are believed to be classically intractable.  A similar approach was codiscovered in the context of simulations for quantum field theories~\cite{Barrett:2013}.  Since this approach was introduced, it has been expanded and enhanced theoretically both in the general sense~\cite{McClean:2015,McClean:2014,Wecker:2015a} and for specific use with ion trap quantum computers~\cite{Yung:2014}.  Recently, variants using a similar approach for thermodynamic properties and extended systems have also appeared~\cite{Dallaire:2015,Bauer:2015}.   Moreover it has been speculated that the robustness and resource adaptive nature of this approach places it as a candidate for one of the first algorithms to surpass the capabilities of a classical computer on a pre-threshold or minimally error corrected quantum device~\cite{Wecker:2015a}.  This conjecture is supported by recent experimental work comparing the performance of quantum phase estimation with variational approaches on superconducting qubits~\cite{OMalley:2015}.  We strengthen this evidence here by showing that it holds in a parameterization-independent model of these variational algorithms in non-ideal conditions.

\begin{figure}[t!]
\includegraphics[width=8.0 cm]{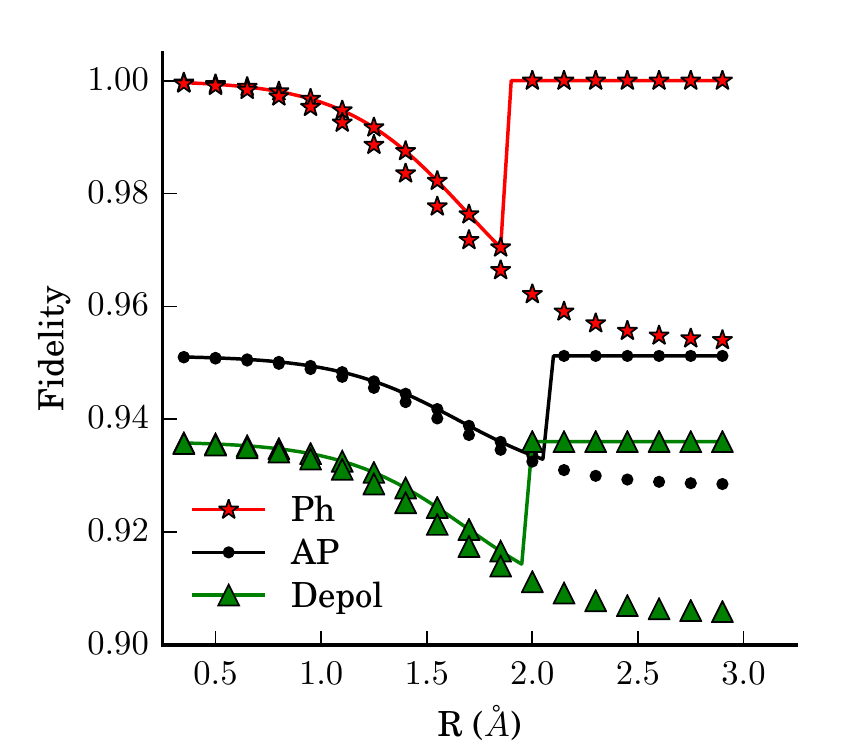}
\caption{The fidelity of the initial and final states after being passed through three different quantum channels (Ph--Dephasing, AP--Amplitude and Phasing Damping, Depol--Depolarizing) improves under variation in the presence of the channel (VCS solution, solid lines with markers) when compared to the exact less ground state as input (ground state of untransformed Hamiltonian, markers only).  The channels are characterized by an experiment time $T_p$ relative to coherence time parameters $T_1=T_2=T_{\text{depol}}$ time such that $T_p/T_1= 0.05$.\label{fig:ChannelFidelities}}
\end{figure}

To understand the performance of a hybrid quantum-classical variational approach on a quantum device experiencing interactions with an environment, here we develop a theoretical model of variational state preparation that we term the variational channel state (VCS) preparation.  Specifically, we define the VCS model as the preparation of an arbitrary pure state followed by the action of a quantum channel defined by a set of Kraus operators~\cite{Kraus:1983}.  The purpose of this model is to allow one to the study the optimal possible performance of quantum-classical variational algorithms in experiments separate from the considerations of ansatz choice or experimental protocol.

In this model, the problem to solve is to find the pure state $\ket{\Psi}$ that minimizes the energy given a target Hamiltonian $H$ after action by a quantum channel that maps $\rho = \ket{\Psi}\bra{\Psi} \rightarrow \sum_i K_i \rho K_i^\dagger$, where $K_i$ are the Kraus operators defining effective non-unitary actions of a dissipative quantum channel, potentially determined by prior experiments.  Mathematically we may state this as choosing the pure state $\ket{\Psi}$ that minimizes
\begin{align}
  \Tr\left[\left(\sum_i K_i\ket{\Psi}\bra{\Psi} K_i^\dagger\right) H\right].
\end{align}
 This problem is equivalent to an eigenvalue problem on the transformed Hamiltonian $H' = \sum_i K_i^\dagger H K_i$ (see supplemental materials for a short proof) such that one may solve
\begin{align}
 H' \ket{\Psi} = E \ket{\Psi}
\end{align}
for the lowest eigenvalue and eigenvector pair to find the solution, which both quantifies the optimal performance of a quantum variational algorithm in these conditions and determines the input state that achieves this optimal performance, independent of state parameterization.

In Fig. \ref{fig:ChannelFidelities} we use the VCS model to compare the fidelity of a 4-qubit quantum state representing H$_2$ communicated through several channels with and without variational optimization, with details of the channels given as supplemental information.  We find that variational optimization in the presence of the channel is able to improve the fidelity and find decoherence resistant subspaces automatically in some cases.  In the case of the dephasing channel, the variational algorithm automatically locates a decoherence free subspace, whereas input of the ideal solution without variational relaxation (or exact diagonalization of the untransformed Hamiltonian, possible in this case due to the limited number of qubits) in the presence of the channel degrades in quality.  The discontinuities in the variational curves correspond to a spin symmetry breaking in the Hamiltonian resulting from an effective interaction induced by the quantum channel.  Thus we see that the variational eigensolver is partially self-correcting in the presence of inevitable qubit decay and dephasing.

We now move on to extensions of the variational method to the capturing of excited states.  To date, hybrid variational quantum-classical algorithms have focused on the ground state in ideal conditions, however we will show that through a straightforward extension of the original machinery, one may both substantially mitigate decoherence and obtain excited states. The original VQE algorithm determines 1- and 2- electron reduced density matrices (1- and 2-RDM) of the system from which static properties of the chemical or material can be determined without any additional quantum experiments.  The 1- and 2- electron reduced density matrices for fermionic systems are defined as
\begin{align}
 {}^{1} D^{i}_{k} &= \bra{\Psi} a_i^\dagger a_k \ket{\Psi} = \Tr[a_i^\dagger a_k \rho] \\
 {}^{2} D^{ij}_{kl} &= \frac{1}{2} \bra{\Psi} a_i^\dagger a_j^\dagger a_l a_k \ket{\Psi} 
 = \frac{1}{2}\Tr[ a_i^\dagger a_j^\dagger a_l a_k \rho]
\end{align}
where here $a_i^\dagger$ and $a_i$ are fermionic creation and annihilation operators acting on spin-orbitals or a generic lattice and $\rho=\ket{\Psi}\bra{\Psi}$ but may represent a more general mixed quantum state $\rho$.   The average energy is obviously expressible as the following contraction once the 1- and 2-RDM are determined
\begin{align}
 \avg{H} = \sum_{ik} h_{ik} {}^{1}D^i_k + \sum_{ijkl} h_{ijkl} {}^{2}D^{ij}_{lk}.
\end{align}

\begin{figure}[t!]
\includegraphics[width=7.0 cm]{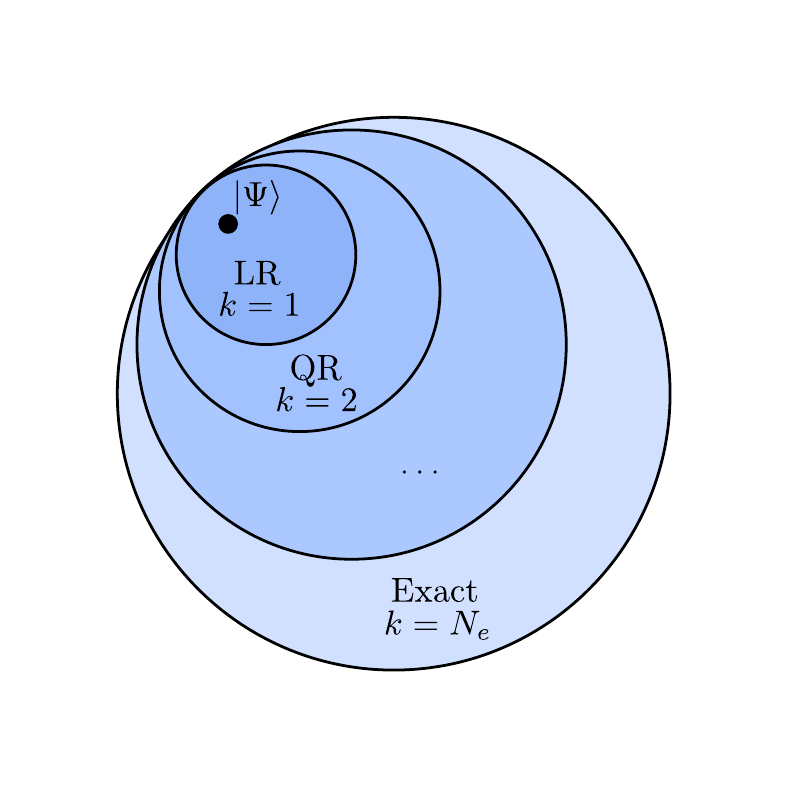}
\caption{A cartoon schematic of the basis hierarchy obtained from expanding about the VQE solution reference.  At $k=1$ one has the linear response (LR) subspace and at $k=2$ one has the the quadratic response  (QR) space continuing to $k=N_e$ where one spans the entire subspace corresponding to the particle number of the reference state. \label{fig:QSEHierarchy}}
\end{figure}

Extending this idea, we now develop a method that requires only a polynomial number of additional measurements to determine the 3- and 4-RDM of the system (see supplemental information for explicit matrix elements), from which excited state energies and properties can be determined.  More explicitly, we expand about the reference $\ket{\Psi}$ to form a linear subspace spanned by the vectors
\begin{align}
 a_i^\dagger a_j \ket{\Psi}.
\end{align}
The justification for including these particular states is that they are the dominant contribution in a linear response (LR) theory of local time-dependent perturbations to the system~\cite{Volker:2014}.  

\begin{figure}[t!]
\includegraphics[width=8.0 cm]{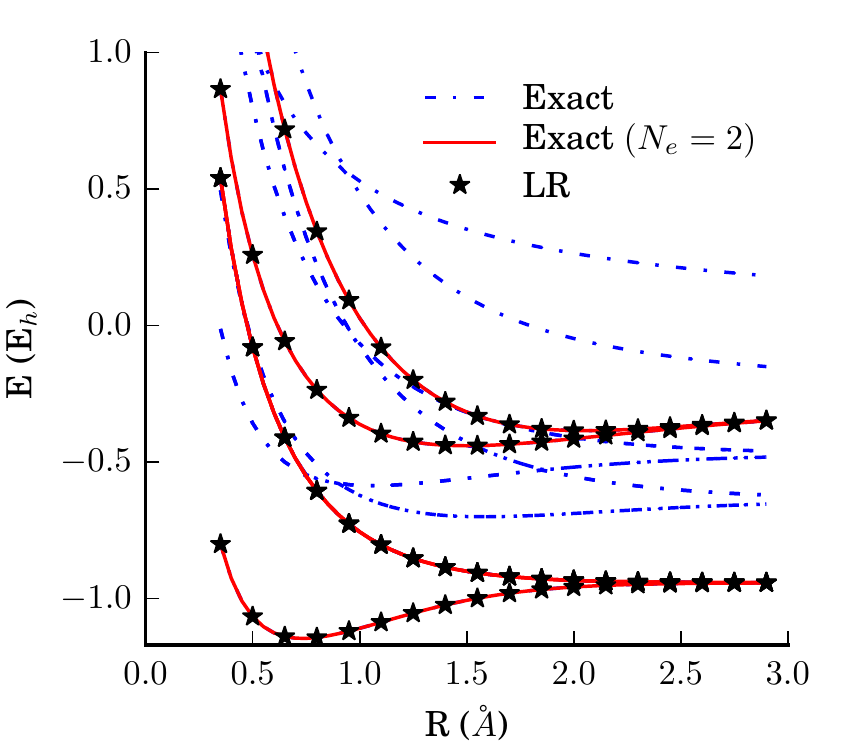}
\caption{The energy of different electronic states as a function of internuclear separation for the H$_2$ molecule in a minimal STO-6G basis, encoded into qubits by the Jordan-Wigner mapping.  Both the entire exact spectrum is shown with dot-dashed lines, as well as the exact spectrum restricted to a neutral molecule $(N_e = 2)$.  In this case the linear response (LR) approximation from an exact reference is sufficient to capture exactly the excited states with the correct number symmetry, as the excitation operators conserve the number from the exact reference state. \label{fig:ExactLRSpectrum}}
\end{figure}

In this linear subspace, the optimal solution within this subspace can be found by solving the generalized eigenvalue problem
\begin{align}
 H_{\text{LR}}C = S_{\text{LR}}CE
\end{align}
for the ground and excited states, where $H_{\text{LR}}$ is the Hamiltonian in this subspace, $S_{\text{LR}}$ is the overlap matrix, $E$ is the diagonal matrix of eigenvalues and $C$ is the matrix of eigenvectors. One may continue to expand the subspace about the reference to an arbitrary order.  This forms a natural hierarchy of subspaces built from the quantum reference state with bases
\begin{align}
 \mathcal{B}_f^k = \{ a_{i_1}^\dagger a_{j_1} a_{i_2}^\dagger a_{j_2} \dots a_{i_k}^\dagger a_{j_k} \ket{\Psi} \ | \ i_k, j_k \in [1,M] \}.
\end{align}
where $\mathcal{B}_f^1$ is clearly the linear response space above, with more and more of the space being spanned until $k=N_e$ and $\mathcal{B}^{N_e}_f$ spans the entire $N_e$-Fermion space. At this point, the classical diagonalization is equivalent to classical exact diagonalization and provides an exact result but has a computational cost that scales exponentially in the size of the system.  While exact diagonalization is not advantageous from a complexity point of view, at fixed levels of the hierarchy before this, one efficiently determines a result that is difficult to obtain classically by virtue of the difficulty of preparing and manipulating $\ket{\Psi}$ and attains more information from $\ket{\Psi}$ from only additional measurements and classical computation.  We refer to this approach generically as fermionic quantum subspace expansion (QSE).  A cartoon schematic of this work is depicted in Fig. \ref{fig:QSESummarySchematic}, where the effect of a quantum channel contracts an ideal pure state, and expanding about the result allows one to compensate for the effect of the dissipative channel while also capturing additional information with the structure of the linear subspace.

We assess the performance of the QSE extension to the original hybrid quantum classical approach on the spectrum of a simple molecule, H$_2$ in a minimal STO-6G basis~\cite{Hehre:1969} under the Jordan-Wigner (JW) qubit encoding~\cite{Jordan1928}, using the VCS model. 

First we examine the performance of the fermionic LR expansion in determining excited states on the exact ground state of H$_2$.  This allows one to understand properties of the method in situations where very good approximations to the ground state may be prepared. The excellent accuracy of the method in this case is exemplified in Fig. \ref{fig:ExactLRSpectrum}.   One sees from this plot a nice feature of the LR method, which is that it confines one to the physical subspace of $N_e = 2$ particles exactly even though the Jordan-Wigner transformation encodes the unphysical space of $N_e=0$ to the number of spin-orbital sites.

\begin{figure}[t!]
\includegraphics[width=8.0 cm]{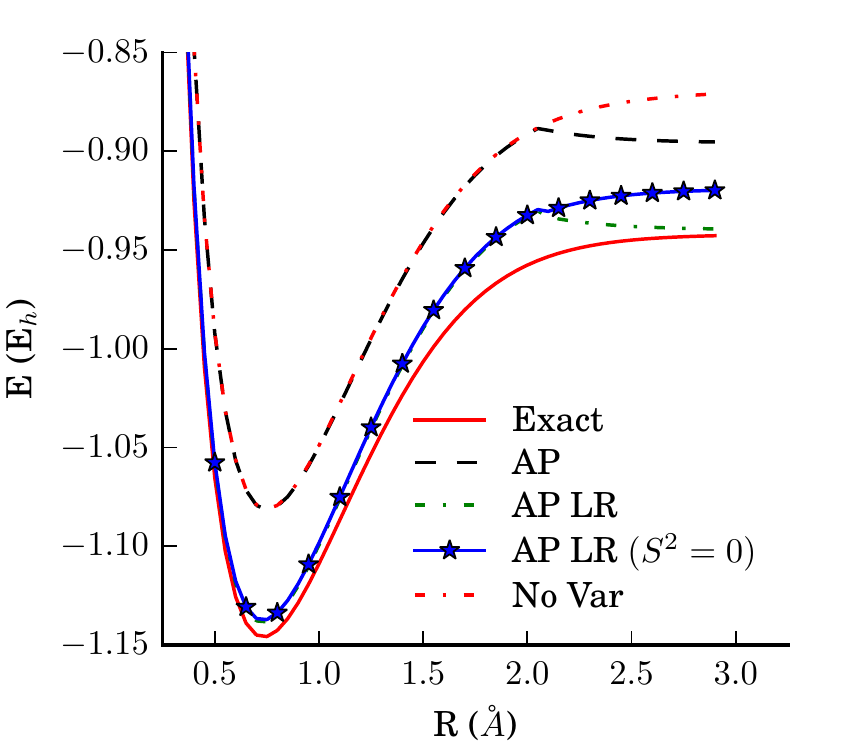}
\caption{In examining the energy as a function of separation for the ground state of H$_2$ under a particular amplitude and phase damping channel (AP), one sees that in the linear response subspace (AP LR), the solution quality in increased with respect to the optimal solution of the quantum channel model (AP).  The qualitative kink in the approximate solution can be repaired by enforcing the correct spin symmetry $(S^2 = 0)$. This also demonstrates the effect on the energy between variational minimization with the channel (AP) and without (No Var). \label{fig:APGroundStateRepairLR}}
\end{figure}

In an imperfect preparation, a quantum channel effectively restricts the space of preparable quantum states.  As such, it's reasonable that re-expanding the resulting state and solving the problem within the expanded space may also help to improve the quality of the ground state.  This calculation is shown in Fig. \ref{fig:APGroundStateRepairLR} for an amplitude and phase damping channel with an experiment time $T_p$ relative to coherence parameters $T_1=T_2$ of $T_p/T_2 = 0.05$, demonstrating the capability of the expansion to improve the quality of the solution under noise.  
n this work, we explore a variational channel state model to understand the performance of quantum variational algorithms in currently-realizable physical systems.  We introduce a simple but powerful approach that naturally mitigates noise, improves estimates of the ground state, and finds excited states based on the projective measurement scheme of the original VQE, which we call the quantum subspace expansion.  Additionally, we develop a variational channel state model (VCS) to understand the potential performance of quantum variational algorithms in non-ideal conditions.  This approach motivates a general hierarchy of quantum-classical approximations, in which a tradeoff between numbers of measurements and system accuracy can be easily tailored to suit computational purposes.  We believe our advances pave the way for better understanding of quantum devices in the role of co-processors and pushes the boundaries of our capabilities closer to the edge of outperforming a purely classical computing device in the near future.

\subsection{Acknowledgements}
The authors acknowledge valuable discussions with Irfan Siddiqi and Chris Macklin.  J.R.M. is supported by the Luis W. Alvarez fellowship in computing sciences at Lawrence Berkeley National Laboratory.  M.E.S. acknowledges support from the Fannie and John Hertz Foundation.  This work was supported by Laboratory Directed Research and Development (LDRD) funding from Berkeley Lab, provided by the Director, Office of Science, of the U.S. Department of Energy under Contract No. DE-AC02-05CH11231. 

\section{Supplemental Information I}
\subsection{Quantum Chemistry}
The electronic structure problem is a problem of great interest due both to its ability to accurately model real molecules from first principles and the potential of quantum computers to greatly accelerate finding its solutions.  Here we provide background on this topic and its mapping to quantum computers for interested readers.  The problem is defined by the electronic eigenstates of a fixed nuclear configuration with positions and charges $R_i$ and $Z_i$ with a fixed number of electrons $N_e$.  Under the Born-Oppenheimer approximation, the non-relativistic Hamiltonian governing the interactions is given by
\begin{align}
H &= - \sum_i \frac{\nabla_{r_i}^2}{2} - \sum_{i,j} \frac{Z_i}{|R_i - r_j|} \notag \\
&+ \sum_{i, j > i} \frac{Z_i Z_j}{|R_i - R_j|} + \sum_{i, j>i} \frac{1}{|r_i - r_j|}
\end{align} in atomic units, and $R_i$ are nuclear positions, $r_i$ electronic positions, and $M_i$ are nuclear masses.  This real-space representation where fermion anti-symmetry may be enforced in the solutions is called the first-quantized representation.  While progress has been made in solving the first-quantized problem on a quantum computer~\cite{Kassal:2008,Ward:2009,Toloui:2013,Welch:2014,Whitfield:2015a}, in this work we will focus on the case where the solution is projected into a finite orthonormal basis and anti-symmetry is enforced through the operators, also known as the second quantized approach~\cite{Helgaker:2014}.  In this approach, the Hamiltonian is given by
\begin{align}
H = \sum_{pq} h_{pq} a^{\dagger}_p a_q + \frac{1}{2} \sum_{pqrs} h_{pqrs} a^{\dagger}_p a^{\dagger}_q a_r a_s
\end{align}
where the coefficients are determined by the integrals over the chosen finite basis as
\begin{align}
h_{pq} &= \int \ d\sigma \ \varphi_p^*(\sigma) \left(\frac{-\nabla_r^2}{2} - \sum_i \frac{Z_i}{|R_i - r|} \right)\varphi_q(\sigma)  \\
h_{pqrs} &= \int \ d\sigma_1 \ d\sigma_2 \frac{\varphi_p^*(\sigma_1)\varphi_q^*(\sigma_2) \varphi_s(\sigma_1)\varphi_r(\sigma_2)}{|r_1 - r_2|}
\end{align}
where $\varphi_i$ are spin-orbitals and $\sigma_i$ are the spatial and spin degrees of freedom of an electron as $\sigma_i = (r_i, s_i)$.  The operators $a_i^\dagger$ and $a_i$ obey the standard fermion commutation relations as
\begin{align}
\{a_p^\dagger, a_r\} &\equiv a_p^\dagger a_r + a_r a_p^\dagger = \delta_{p,r} \\
\{a_p^\dagger, a_r^\dagger \} &= \{a_p, a_r\} = 0.
\end{align}

In quantum computing, one must represent anti-symmetric fermions by distinguishable qubits.  At least two isomorphisms are known for accomplishing this, namely the Jordan-Wigner~\cite{Jordan1928,Somma2002} and Bravyi-Kitaev transformations~\cite{Bravyi2000,Seeley2012,Tranter:2015}.  Each of these approaches have their tradeoffs in implementation, but in this work we will use the Jordan-Wigner transformation defined by
\begin{align}
a_p^\dagger  &= ( {\prod\nolimits_{m < p} {\sigma _m^z} } ) \sigma _p^+ \\
a_p &= ( {\prod\nolimits_{m < p} {\sigma _m^z} } )\sigma _p^- \\
\sigma^\pm &\equiv \left( {{\sigma ^x} \mp i{\sigma ^y}} \right)/2
\end{align}
This encoding allows one to express the second-quantized Hamiltonian entirely in terms of tensor products of Pauli operators.  Moreover, this transformation leaves the number of terms the same up to a constant factor, and may be used to derive the Pauli representation of any desired fermion operator.  

One practical property to note with regards to this mapping is that it encodes all particle number manifolds.  That is, the quantum chemistry Hamiltonian commutes with the number operator $N=\sum_i a_i^\dagger a_i$, such that the number of electrons is a good quantum number. Equivalently, the Hamiltonian can be decomposed into a block diagonal representation where different number states are uncoupled.  Using this symmetry classically allows one to remain in the desired number manifold at all times.  However, on a quantum device, the plethora of unphysical excited states can pollute the spectrum as a result of this wasteful encoding.  While some approaches have been developed to project out only the correct states at the operator level~\cite{Moll:2015}, provably polynomial methods for doing this are still under development.  We explore in the body of this work how these unphysical excited states may enter in practice, and show how they can be tempered using the extra structure in our method.

\subsection{Hybrid Quantum-Classical Variational Approach}
Quantum phase estimation provided the first demonstration that quantum computers could aid in the solution of electronic structure problems for quantum chemistry~\cite{Aspuru:2005}.  However, this approach requires long coherent sequences of quantum operations that are not easily implemented on current quantum architectures.  In order to study this problem on current and near-future architectures, a hybrid quantum-classical approach called the variational quantum eigensolver (VQE) was developed, which leverages classical computing power alongside the power of a quantum device to minimize coherence time requirements. Here we briefly review the parts of this algorithm relevant to the main body of this work, and refer readers to the original works for more detailed algorithmic analysis of the original approach~\cite{Peruzzo2014,McClean:2015}.

The VQE approach depends on the choice of a state ansatz parameterized on some set of experimental parameters $\vec{\theta}$.  These parameters could be used to specify a gate sequence such as in the unitary coupled cluster or parameterized adiabatic state preparation approach~\cite{McClean:2015,OMalley:2015}, or they could be more directly related to the hardware such as the angles on beamsplitters as was used in the first experimental implementation of the algorithm~\cite{Peruzzo2014}.  In either case, the state that is produced becomes a function of the discrete set of input parameters, and we may call the resulting state $\ket{\Psi(\vec{\theta})}$.  The goal of the algorithm is to find a set of parameters $\vec{\theta}$ such that the expectation value of the energy $\avg{H}$ is a minimum.  That is, we exploit the Rayleigh-Ritz variational formulation of the eigenvalue problem~\cite{Weinstein:1934,MacDonald:1934} such that the best approximation to the ground state eigenvalue may be found from
\begin{align}
 \min_{\vec{\theta}} \avg{H}(\vec{\theta}) = \frac{\bra{\Psi(\vec{\theta})} H \ket{\Psi(\vec{\theta})}}{\braket{\Psi(\vec{\theta})}{\Psi(\vec{\theta})}}.
\end{align}
Generically, the VQE approach can be broken into three subtasks, namely preparation of $\ket{\Psi(\vec{\theta})}$, measurement of $\avg{H}(\vec{\theta})$ with respect to $\ket{\Psi(\vec{\theta})}$, and the update of $\vec{\theta}$ based on the measured values.  In this work we focus on how projective measurement type approaches can be extended and better utilized.

In particular, we advocate a projective measurement approach for the determination of the average energy $\avg{H}(\vec{\theta})$ through repeated state preparation and partial tomography.  The specific Pauli measurements one performs following state preparation can be derived from the mapping from fermionic operators to qubits such as the JW transformation.  That is, without considering potential variance reducing optimizations, the estimator for our average may be constructed as
\begin{align}
 \avg{H}(\vec{\theta}) = \sum_{ij} \avg{a_i^\dagger a_j}(\vec{\theta}) + \frac{1}{2}\sum_{ijkl} h_{ijkl} \avg{a_i^\dagger a_j^\dagger a_k a_l}(\vec{\theta})
\end{align}
with each average $\avg{a_i^\dagger a_j}(\vec{\theta})$ and $\avg{a_i^\dagger a_j^\dagger a_k a_l}(\vec{\theta})$ being determined by first mapping the operator to a Pauli string through the JW transformation, and determining the average by repeated state preparation and measurement of the resulting term on the quantum state $\ket{\Psi(\vec{\theta})}$.  The energy estimator is then evaluated by classically adding each of the individual estimators along with the weight factors $h_{ijkl}$.  However, as we will  emphasize, the information gained by evaluating the expectation values $\avg{a_i^\dagger a_j^\dagger a_k a_l}$ on a quantum state is actually far greater than simply the energy.

As has been shown previously~\cite{McClean:2015}, the minimization can be modified using penalty terms to enforce certain constraints on the final solution, similar to other penalty methods used in quantum computing~\cite{Bookatz:2014}.  This is done by modifying to Hamiltonian to
\begin{align}
 H \rightarrow H + \sum_i \lambda_i (O_i - o_i)^2
\end{align}
where $O_i$ and $o_i$ are the corresponding symmetry operators and eigenvalues desired.  In the limit that the penalty parameters $\lambda_i$ approach infinity, the solutions of the minimization exactly satisfy the desired symmetry, assuming it is possible with the given parameterization of the wavefunction.  In practice finite values of $\lambda_i$ are usually sufficient to satisfy the symmetry to a desired precision.  Of particular interest in this work will be the spin and number operators $S^2$ and $N$.  We note that if the symmetry operator is a one-fermion operator such as the number operator $N$, then this modification requires no additional measurements beyond those required for the original measurements.

\subsection{Symmetries in the subspace}
One advantage of the QSE approach is that the additional structure of the linear subspace allows one to exactly enforce symmetries.  As discussed in the body of the text, with this representation of the operator and overlap in this linear subspace, the optimal solution within this subspace can be found by solving the generalized eigenvalue problem
\begin{align}
 H_{\text{LR}}C = S_{\text{LR}}CE.
\end{align}

Expanding the problem into a linear subspace also allows the use of additional analysis and solution tools.  One tool of great practical is the ability to enforce particular symmetries in this linear subspace.  
For example, in the JW encoding of the quantum chemistry Hamiltonian, all number states from $N_e = 0$ to $M$ are encoded, however often only a particular number state is of physical interest.  The non-linear penalty method introduced for the VQE is one way to enforce this symmetry on the reference, however it can be prohibitively expensive and also may not generalize well to excited states if they are of a different symmetry than the ground state.  An example of this is when the ground state is known to be a spin singlet while excited states of interest are spin triplets.  

To enforce desired symmetries in the linear subspace, one first constructs the matrix representation of both the Hamiltonian and the symmetry operator $O$ in the linear response subspace as was done for the Hamiltonian.  General expressions for these expansions are given later in the supplemental information.  The eigenvectors corresponding to the desired symmetry eigenvalues of $O_{\text{LR}}$ may then be used to project the Hamiltonian into a particular symmetry subspace, where a subsequent diagonalization yields the optimal solution subject to the symmetry constraint.  

\subsection{Quantum Channel Model Solution}
Here we present the short proof that the quantum channel state preparation model is equivalent to a Hermitian eigenvalue problem on the transformed Hamiltonian $H'=\sum_i K_i^\dagger H K_i$.  Starting with the original problem
\begin{align}
  \min_{\ket{\Psi}} \Tr\left[\left(\sum_i K_i \ket{\Psi}\bra{\Psi} K_i^\dagger\right) H\right].
\end{align}
we require that the function we are minimizing vanish under arbitrary variations in the state $\bra{\Psi} \rightarrow \bra{\Psi} + \bra{\delta \Psi}$ (note that we only need consider variations in the bra(dual) for simplicity due to the symmetric real valued nature of this functional), and enforce the constraint of normalization on the pure state through a Lagrange multiplier $E$, resulting in
\begin{align}
\Tr\left[\left(\sum_i K_i \ket{\Psi}\bra{\delta \Psi} K_i^\dagger\right) H\right] - E \braket{\delta \Psi}{\Psi} = 0
\end{align}
By cyclic invariance of the trace and independence of $\ket{\Psi}$ from $i$, equivalently
\begin{align}
\Tr \left[ \ket{\Psi}\bra{\delta \Psi} H' \right] - E \braket{\delta \Psi}{\Psi} = 0
\end{align}
Expanding the trace over a basis composed of $\ket{\Psi}$ and its orthogonal complement 
\begin{align}
\bra{\delta \Psi} H' \ket{\Psi} - E \braket{\delta \Psi}{\Psi} = 0.
\end{align}
By requiring that this vanish under arbitrary variations $\bra{\delta \Psi}$ we arrive at the eigenvalue equation
\begin{align}
 H' \ket{\Psi} = E \ket{\Psi}
\end{align}
and the Hermiticity of $H'$ follows trivially from the Hermiticity of $H$ and the form of $H'$, guaranteeing it may be diagonalized by a unitary matrix.

\subsection{Quantum Channels}
While the quantum channels used in this work are standard, for completeness we detail the specific Kraus operators and channels used in this section as well as our mappings between the experimental parameters corresponding to the total experiment time $T_p$, the decay time $T_1$, and the dephasing time $T_2$.  In particular, we will recall the Kraus operator definitions for the dephasing, amplitude and phase damping, and depolarizing channel in terms of these parameters.  

One of the simplest quantum channels is the dephasing channel, which is related to the T$_2$ time of quantum systems.  It has a set of Kraus operators defined by
\begin{align}
F_P(\tilde p_i)[\rho] &= \sum_i K_i \rho K_i^{\dagger} \\
K_0 &= \sqrt{1.0 - \frac{\tilde p_i}{2}} \ I \\
K_1 &= \sqrt{\frac{\tilde  p_i}{2}} \ Z
\end{align}
where $Z$ is the standard Pauli z matrix. 
The effect of this map on an arbitrary one-particle density matrix is given by
\begin{align}
F_{P}(\tilde p_i)[\rho] &= \left( \begin{array}{cc} \rho_{00} &  (1-\tilde p_i) \rho_{01} \\ (1-\tilde p_i)  \rho_{10} & \rho_{11} \end{array} \right)
\end{align}
where we choose the values of $\tilde p_i=1-\exp(-T_p/T_2)$ such that the resulting action on a one qubit density matrix is given by
\begin{align}
F_{P}(\tilde p_i)[\rho] &= \left( \begin{array}{cc} \rho_{00} &  e^{-T_p/T_2} \rho_{01} \\ e^{-T_p/T_2}  \rho_{10} & \rho_{11} \end{array} \right)
\end{align}
Another important quantum channel we will consider in more detail in this work is an amplitude and phase damping channel applied independently to each qubit with three input parameters, namely a total time of state preparation T$_p$ and the qubit decay and dephasing times T$_1$ and T$_2$.  
Mathematically, we construct the amplitude and phase damping channels in a Kraus operators formalism such that the quantum map $F_{AP}(p_i)[\rho] = F_{P}(\tilde p_i) \left[F_{A}(p_i)[\rho] \right]$ where $F_A$ and $F_P$ are amplitude and phase damping operators.  $F_P$ is defined as above, and $F_A$ is given by
\begin{align}
F_A(p_i)[\rho] &= \sum_i K_i \rho K_i^{\dagger} \\
K_0 &= \left( \begin{array}{cc} 1 & 0 \\ 0 & \sqrt{1 - p_i} \end{array} \right) \\
K_1 &= \left( \begin{array}{cc} 0 & \sqrt{p_i} \\ 0 & 0 \end{array} \right).
\end{align}
The probabilities are determined by the probability such an event would have occurred in the preceding gate operation, given some values of $T_1$ and $T_2$.

The effect of the composite map on an arbitrary one-particle density matrix is given by
\begin{align}
F_{AP}(p_i)[\rho] &= \left( \begin{array}{cc} \rho_{00} + p_i \rho_{11} &  (1-\tilde p_i) \sqrt{1-p_i} \rho_{01} \\ (1-\tilde p_i) \sqrt{1-p_i}  \rho_{10} & (1-p_i)\rho_{11} \end{array} \right)
\end{align}
and the values of $p_i$ and $\tilde p_i$ are determined such that
\begin{align}
F_{AP}(\tilde p_i)[\rho] &=
\left( \begin{array}{cc} \rho_{00} + (1 - e^{-T_p/T_1}) \rho_{11} & e^{-T_p/T_2} \rho_{01} \\ e^{-T_p/T_2} \rho_{10} & e^{-T_p/T_1} \rho_{11} \end{array} \right).
\end{align}
It's clear from this construction that the relevant dimensionless parameters that determine performance will be $T_p/T_1$ and $T_p/T_2$, or the ratios of the state preparation time to the decay and dephasing time of the qubits.

Finally, we also consider the depolarizing quantum channel $F_D$ that corresponds to uniform contraction of the Bloch sphere of a qubit, and has corresponding Kraus operators given by
\begin{align}
F_D(p_i)[\rho] &= \sum_i K_i \rho K_i^{\dagger} \\
K_0 &= \sqrt{1-p_i} \\
K_1 &= \sqrt{\frac{p_i}{3}} X \\
K_2 &= \sqrt{\frac{p_i}{3}} Y \\
K_3 &= \sqrt{\frac{p_i}{3}} Z
\end{align}
where $X$, $Y$, and $Z$ correspond to the standard Pauli matrices.  In the case of the depolarizing channel, we choose $p_i=1-\exp(-T_p/T_2)$.

\begin{figure}[t!]
\includegraphics[width=8.0 cm]{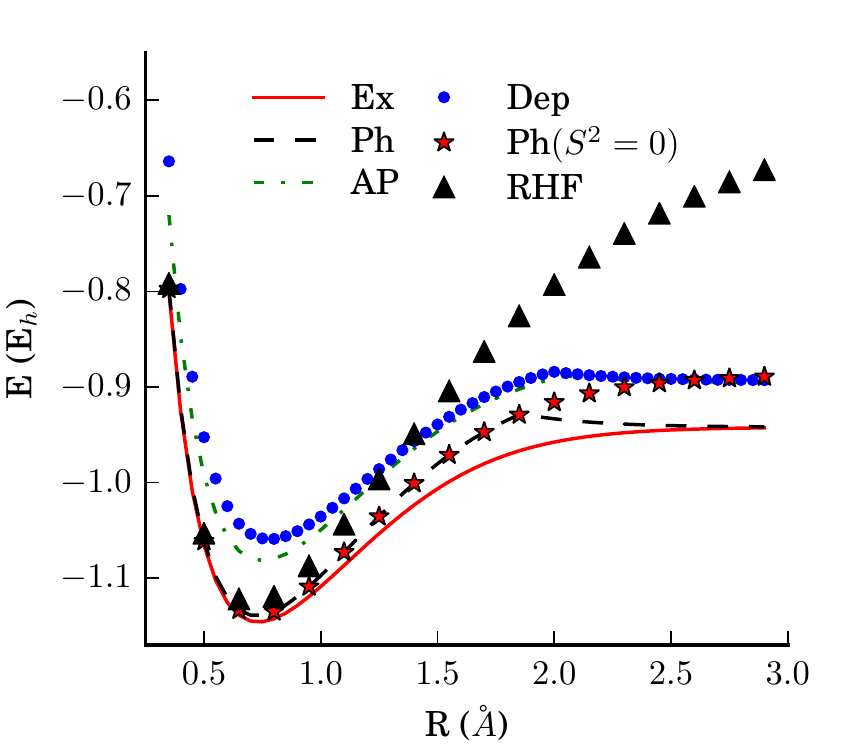}
\caption{The exact solution of the VCS model for the ground state of H$_2$ is shown for a number of different quantum channels including amplitude and phase damping (AP), dephasing only (Ph), and depolarizing noise (Dep). These results are shown along side both the exact (Ex) ground state, an anti-symmetric product state approximation (RHF), and the best solution of a dephasing channel constrained to have correct spin $(S^2 = 0)$.  It is seen that dephasing noise is sufficient to destroy the entanglement required to describe the dissociated limit, in that the solution without symmetry constraints obtains an accurate energy, but only by breaking spin symmetry in an unphysical way, as evidenced by the difference when compared to the optimal dephasing solution under symmetry constraints.  Other types of noise raise the energy of the whole curve due to number symmetry breaking.  The kink in the curves without symmetry enforcement results from a spin symmetry breaking in the variationally optimal solution in the presence of decohering noise.  \label{fig:QCGroundState}}
\end{figure}

The qualitative effects of a different number of channels under the VCS model on the electronic ground state of H$_2$ in a STO-3G basis are depicted in Fig. \ref{fig:QCGroundState}.  In this work, all channels utilize parameters of $(T_p/T_1) = (T_p/T_2) = 0.05$, or a total gate sequence time corresponding to roughly 5\% of an expected coherence time.


\subsection{Spin Quantum Subspace Expansion}
While the fermionic specification of the quantum subspace expansion (QSE) is of great interest for the study chemistry and materials, it is valuable to consider such expansions at the level of qubits as well.  Moreover, before using a quantum device, fermionic problems are first mapped to qubit systems where similar considerations will apply.  Given a quantum state of $N$ qubits $\ket{\Psi}$, we can define a qubit QSE about that state as
\begin{align}
 \mathcal{B}_q^k = \left\{ \sigma_{i_1}^{\alpha_1} \sigma_{i_2}^{\alpha_2} \dots \sigma_{i_k}^{\alpha_k} \ket{\Psi} \ | \ \alpha_i \in [I,X,Y,Z] \right\}
\end{align}
where an operator $\sigma_i^\alpha$ is Pauli operator acting on qubit $i$ and $\alpha$ identifies if the operator is $I$, $X$, $Y$, or $Z$.  This hierarchy expands in a basis that has low Hamming distance (or number of spins different) from the original state.  Whether this constitutes a good approximation hierarchy will depend on this problem of interest, and indeed hierarchies should be based on the interactions the problems are likely to experience.  However, low order truncations of this hierarchy play an interesting role with respect to error suppression.

Specifically, imagine that after preparation of $\ket{\Psi}$ the state is passed through a channel in which one if its spins is acted upon by a Pauli error operator such as $X_1$.  By expanding in $\mathcal{B}^1_q$, the original desired state is contained within the subspace, and the error can be corrected exactly through the solution of the linear eigenvalue problem on $\mathcal{B}^1_q$.  More generically, $k$ qubit errors can be mitigated by solving the problem in the subspace $\mathcal{B}_q^k$, requiring again, only additional measurements and classical computation. This is especially appealing for pre-threshold devices and those with minimal error correction, as it utilizes classical computation to extend the capabilities of the quantum device.  Our numerical work exclusively focuses on the performance of error supression in fermionic QSE in this work, leaving more general expansions as a subject of future research. 

\subsection{Linear Response Representations from RDMs}
In this short section we explicitly construct the representations of one- and two-body fermion operators in the linear response subspace from the reduced density matrices of the system.  The reduced density matrices are defined as
\begin{align}
{}^{k}D^{i_1 i_2 ... i_k}_{j_1 j_2 ... j_k} &= \frac{1}{k!} \bra{\Psi} a_{i_1}^\dagger a_{i_2}^\dagger ... a_{i_k}^\dagger a_{j_k} a_{j_{k-1}}...a_{j_1} \ket{\Psi} \notag \\
&= \frac{1}{k!} \Tr[a_{i_1}^\dagger a_{i_2}^\dagger ... a_{i_k}^\dagger a_{j_k} a_{j_{k-1}}...a_{j_1} \rho]
\end{align}
where we call ${}^kD$ the $k$ fermion reduced density matrix or $k-$RDM.  We will examine matrix elements that couple the reference state $\ket{\Psi}$ denoted by index $g$ and the linear response space.  For any operator $O$, these matrix elements are defined as
\begin{align}
O^{ij}_g &= \bra{\Psi} (a_i^\dagger a_j)^{\dagger} O \ket{\Psi} = \Tr [ (a_i^\dagger a_j)^{\dagger} O \rho] \\
O^{ij}_{kl} &= \bra{\Psi} (a_i^\dagger a_j)^{\dagger} O a_k^\dagger a_l\ket{\Psi} 
= \Tr [ (a_i^\dagger a_j)^{\dagger} O a_k^\dagger a_l \rho]
\end{align}
A crucial factor in all the above calculations is the overlap operator or metric $S$, which in the linear subspace of $\rho$ is given by
\begin{align}
S^{ij}_g &= {}^1D^j_i\\
S^{ij}_{kl} &= \delta_{i k} {}^{1}D^{j}_{l} - 2{\ }^{2}D^{jk}_{li}.
\end{align}
One-electron operators $F = \sum_{pr} a_p^\dagger a_r$ have the following matrix elements
\begin{align}
F^{ij}_g &= \sum_{pr} \left[ \delta_{i p} {}^{1}D^{j}_{r} - 2{\ }^{2}D^{jp}_{ri}\right] \\
F^{ij}_{kl} &= \sum_{pr} \left[ - 2\delta_{i k} {}^{2}D^{jp}_{rl} + \delta_{i p} \delta_{k r} {}^{1}D^{j}_{l} + 2 \delta_{i p} {}^{2}D^{jk}_{rl}  \right. \notag \\
& \left. - 2 \delta_{k r} {}^{2}D^{jp}_{li} - 6{\ }^{3}D^{jkp}_{rli} \right].
\end{align}
Two-body operators $V=\sum_{pqrs} a_p^\dagger a_q^\dagger a_r a_s$ have matrix elements given by
\begin{align}
V^{ij}_g &= \sum_{pqrs} \left[ 2 \delta_{i p} {}^{2}D^{jq}_{sr} - 2 \delta_{i q} {}^{2}D^{jp}_{sr} + 6{\ }^{3}D^{jpq}_{sri} \right] \\
V^{ij}_{kl} &= \sum_{pqrs} \left[ 6 \delta_{i k} {}^{3}D^{jpq}_{srl} + 2 \delta_{i p} \delta_{k r} {}^{2}D^{jq}_{sl} - 2 \delta_{i p} \delta_{k s} {}^{2}D^{jq}_{rl} \right. \notag \\
&- 6 \delta_{i p} {}^{3}D^{jkq}_{srl} - 2 \delta_{i q} \delta_{k r} {}^{2}D^{jp}_{sl} + 2 \delta_{i q} \delta_{k s} {}^{2}D^{jp}_{rl} \notag \\
&+ 6 \delta_{i q} {}^{3}D^{jkp}_{srl} + 6 \delta_{k r} {}^{3}D^{jpq}_{sli} - 6 \delta_{k s} {}^{3}D^{jpq}_{rli} \notag \\
& \left. - 24{\ }^{4}D^{jkpq}_{srli} \right].
\end{align}
\begin{figure}[t]
\includegraphics[width=8.0 cm]{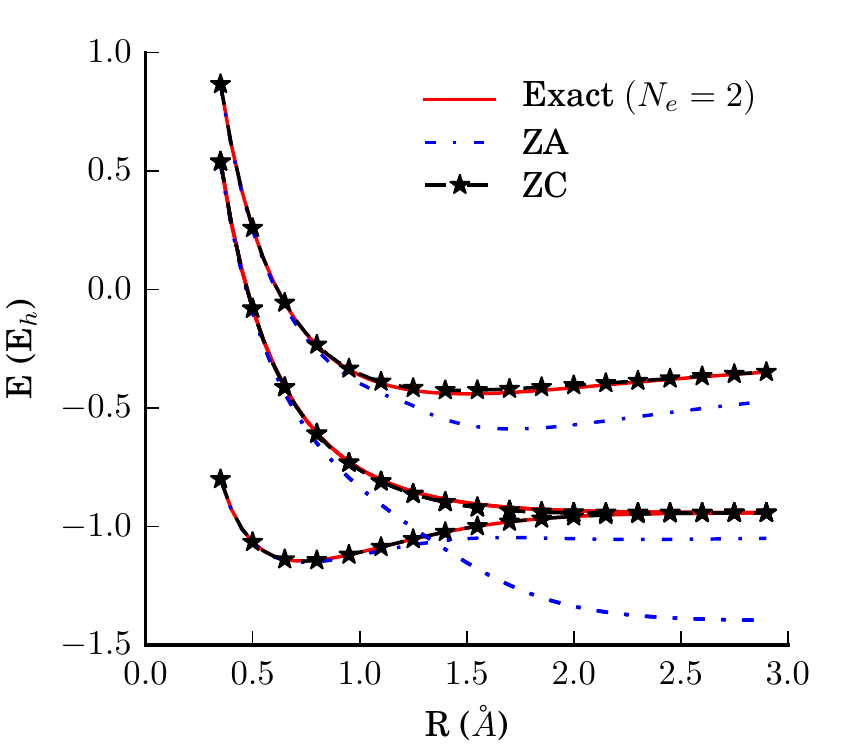}
\caption{The ground and first two excited states of H$_2$ are plotted as a function of internuclear separation using both the Zero Approximation (ZA) and the Zero Approximation under the commutator construction (ZC).  Both methods require only the original measurements used for the ground state to approximate the excited states, and we see here that ZC achieves an extremely high accuracy, while ZA is qualitatively correct in some cases but produces sub-variational solutions in others.
\label{fig:ZeroApprox}}
\end{figure}

The Hamiltonian and any other operators expressed as sums of one- and two-body operators in the linear response subspace can be formed by simply summing these expressions together.
\subsection{Cumulant Expansions of the $k-$RDMs}
Here we document the cumulant expansions of the reduced density matrices up to $k=4$, which are important in approximation schemes for the reduced density matrices.  The fermionic $k-$RDM on a quantum state $\rho$ is defined by
\begin{align}
 {}^{k}D^{i_1 i_2...i_k}_{j_1 j_2 ... j_k} = \frac{1}{k!}\Tr \left[ a_{i_1}^\dagger a_{i_2}^\dagger ... a_{i_k}^\dagger a_{j_k} a_{j_k-1} ... a_{j_1} \rho \right]
\end{align}
The cumulant expansions decompose the reduced density matrices into their non-separable (connected) components and separable unconnected components, and are quite useful for both developing approximations and enhancing understanding. A convenient notation for expressing these expansions is given by the Grassmann wedge product defined generally by
\begin{align}
 a \wedge b = \left( \frac{1}{N!} \right)^2 \sum_{\pi, \sigma} \epsilon(\pi) \epsilon(\sigma) \ \pi \  \sigma \ a \otimes b
\end{align}
where $\pi$ and $\sigma$ are permutations on the upper and lower indices of the tensor $a\otimes b$ and $\epsilon$ denotes the parity of each permutation.  As an example one might consider the wedge product of a cumulant matrix with itself
\begin{align}
 \left[{}^1\Delta \wedge {}^1 \Delta\right]^{i_1 i_2}_{j_1 j_2} = \frac{1}{2} \left( {}^1\Delta^{i_1}_{j_1} {}^1\Delta^{i_2}_{j_2} - {}^1\Delta^{i_1}_{j_2} {}^1\Delta^{i_2}_{j_1} \right).
\end{align}
With this notation, the reduced density matrices up to $k=4$ are iteratively defined in terms of the cumulant expansions as
\begin{align}
 {}^{1}D &= {}^{1}\Delta \\
 {}^{2}D &= {}^{2}\Delta + {}^{1}\Delta \wedge {}^{1}\Delta \\
 {}^{3}D &= {}^{3}\Delta + 3 {}^{2}\Delta \wedge {}^{1}\Delta + {}^{1}\Delta \wedge {}^{1}\Delta \wedge {}^{1}\Delta \\
 {}^{4}D &= {}^{4}\Delta + 4 {}^{3}\Delta \wedge {}^{1}\Delta + 3 {}^{2}\Delta \wedge {}^{2}\Delta \notag \\
 &+ 6 {}^{2}\Delta \wedge {}^{1}\Delta \wedge {}^{1}\Delta
  + {}^{1}\Delta \wedge {}^{1}\Delta \wedge {}^{1}\Delta \wedge {}^{1}\Delta.
\end{align}
Physically, we may interpret terms such as ${}^{m}\Delta \wedge {}^{n}\Delta$ as the product between irreducible $m$ and $n$ body correlations which contribute to the overall $(m+n)$ body correlations.

\subsection{Approximate QSE}
While the fermionic QSE approach has favorable coherence time requirements, it can still be prohibitively expensive in the number of measurements required.  As such, we examine here some techniques that have been developed in classical electronic structure theory for 2-RDMs for approximating the eigenvalue problem requiring only the 3-RDM or even only the 2-RDM for approximating the solution in the linear response space $B^1_f$ following closely the techniques developed by Mazziotti~\cite{Mazziotti:2003,Greenman:2008}. In such a scheme, even the original measurements used to determine the ground state energy are sufficient to determine approximations for the excited states of the system.  

Suppose that the quantum state prepared $\ket{\Psi}$ is the exact ground state.  In this case, the following commutator identity may be used to remove dependence of the solution on the 4-RDM
\begin{align}
 H^{ij}_{kl} = \bra{\Psi} (a_i^\dagger a_j)^\dagger [H, a_k^\dagger a_l] \ket{\Psi} 
 + E_g \bra{\Psi} (a_i^\dagger a_j)^\dagger a_k^\dagger a_l \ket{\Psi}
\end{align}
where $E_g$ is the eigenvalue associated with the exact ground state, or expectation value of $\ket{\Psi}$ in the case of an approximation, which depends at most on the 2-RDM.  The commutator reduces the rank of this expression such that it depends at most on the 3-RDM requiring only $O(M^6)$ terms to be measured, and is exact in the case that $\ket{\Psi}$ is exact.  The explicit dependence on the 3-RDM can be removed through approximate density matrix reconstruction techniques, requiring only the original 2-RDM measurements to produce excited state approximations.  The simplest such approximation neglects the irreducible 3-body correlations in the commutator expansion above, setting ${ }^3\Delta=0$, where ${ }^3 \Delta$ is the 3 particle cumulant, and we call this approximation the zero in commutator approximation (ZC).

Another similar approximation starts from the original expression for the 4-RDM without the commutator reduction, and assumes that both the irreducible 3 and 4 particle correlations are negligible and reconstructs the 4-RDM from only the 2-RDM assuming ${ }^4\Delta={ }^3\Delta=0$.  We term this this full zero approximation, or ZA. The performance of the ZC and ZA methods are shown in Fig. \ref{fig:ZeroApprox}.  We see that the extra structure in the ZC approximations yields superior qualitative and quantitive accuracy for the sample problem.

\bibliographystyle{apsrev4-1}
\bibliography{QCHierarchyPaper}

\begin{thebibliography}{46}%
\makeatletter
\providecommand \@ifxundefined [1]{%
 \@ifx{#1\undefined}
}%
\providecommand \@ifnum [1]{%
 \ifnum #1\expandafter \@firstoftwo
 \else \expandafter \@secondoftwo
 \fi
}%
\providecommand \@ifx [1]{%
 \ifx #1\expandafter \@firstoftwo
 \else \expandafter \@secondoftwo
 \fi
}%
\providecommand \natexlab [1]{#1}%
\providecommand \enquote  [1]{``#1''}%
\providecommand \bibnamefont  [1]{#1}%
\providecommand \bibfnamefont [1]{#1}%
\providecommand \citenamefont [1]{#1}%
\providecommand \href@noop [0]{\@secondoftwo}%
\providecommand \href [0]{\begingroup \@sanitize@url \@href}%
\providecommand \@href[1]{\@@startlink{#1}\@@href}%
\providecommand \@@href[1]{\endgroup#1\@@endlink}%
\providecommand \@sanitize@url [0]{\catcode `\\12\catcode `\$12\catcode
  `\&12\catcode `\#12\catcode `\^12\catcode `\_12\catcode `\%12\relax}%
\providecommand \@@startlink[1]{}%
\providecommand \@@endlink[0]{}%
\providecommand \url  [0]{\begingroup\@sanitize@url \@url }%
\providecommand \@url [1]{\endgroup\@href {#1}{\urlprefix }}%
\providecommand \urlprefix  [0]{URL }%
\providecommand \Eprint [0]{\href }%
\providecommand \doibase [0]{http://dx.doi.org/}%
\providecommand \selectlanguage [0]{\@gobble}%
\providecommand \bibinfo  [0]{\@secondoftwo}%
\providecommand \bibfield  [0]{\@secondoftwo}%
\providecommand \translation [1]{[#1]}%
\providecommand \BibitemOpen [0]{}%
\providecommand \bibitemStop [0]{}%
\providecommand \bibitemNoStop [0]{.\EOS\space}%
\providecommand \EOS [0]{\spacefactor3000\relax}%
\providecommand \BibitemShut  [1]{\csname bibitem#1\endcsname}%
\let\auto@bib@innerbib\@empty
\bibitem [{\citenamefont {Feynman}(1982)}]{Feynman1982}%
  \BibitemOpen
  \bibfield  {author} {\bibinfo {author} {\bibfnamefont {R.~P.}\ \bibnamefont
  {Feynman}},\ }\href {http://link.aps.org/doi/10.1103/PhysRevLett.108.120603}
  {\bibfield  {journal} {\bibinfo  {journal} {Int. J. Theor. Phys.}\ }\textbf
  {\bibinfo {volume} {21}},\ \bibinfo {pages} {467} (\bibinfo {year}
  {1982})}\BibitemShut {NoStop}%
\bibitem [{\citenamefont {Aspuru-Guzik}\ \emph {et~al.}(2005)\citenamefont
  {Aspuru-Guzik}, \citenamefont {Dutoi}, \citenamefont {Love},\ and\
  \citenamefont {Head-Gordon}}]{Aspuru:2005}%
  \BibitemOpen
  \bibfield  {author} {\bibinfo {author} {\bibfnamefont {A.}~\bibnamefont
  {Aspuru-Guzik}}, \bibinfo {author} {\bibfnamefont {A.~D.}\ \bibnamefont
  {Dutoi}}, \bibinfo {author} {\bibfnamefont {P.~J.}\ \bibnamefont {Love}}, \
  and\ \bibinfo {author} {\bibfnamefont {M.}~\bibnamefont {Head-Gordon}},\
  }\href {http://www.sciencemag.org/content/309/5741/1704.abstract} {\bibfield
  {journal} {\bibinfo  {journal} {Science}\ }\textbf {\bibinfo {volume}
  {309}},\ \bibinfo {pages} {1704} (\bibinfo {year} {2005})}\BibitemShut
  {NoStop}%
\bibitem [{\citenamefont {Kassal}\ \emph {et~al.}(2011)\citenamefont {Kassal},
  \citenamefont {Whitfield}, \citenamefont {Perdomo-Ortiz}, \citenamefont
  {Yung},\ and\ \citenamefont {Aspuru-Guzik}}]{Kassal:2011}%
  \BibitemOpen
  \bibfield  {author} {\bibinfo {author} {\bibfnamefont {I.}~\bibnamefont
  {Kassal}}, \bibinfo {author} {\bibfnamefont {J.~D.}\ \bibnamefont
  {Whitfield}}, \bibinfo {author} {\bibfnamefont {A.}~\bibnamefont
  {Perdomo-Ortiz}}, \bibinfo {author} {\bibfnamefont {M.-H.}\ \bibnamefont
  {Yung}}, \ and\ \bibinfo {author} {\bibfnamefont {A.}~\bibnamefont
  {Aspuru-Guzik}},\ }\href@noop {} {\bibfield  {journal} {\bibinfo  {journal}
  {Ann. Rev. Phys. Chem.}\ }\textbf {\bibinfo {volume} {62}},\ \bibinfo {pages}
  {185} (\bibinfo {year} {2011})}\BibitemShut {NoStop}%
\bibitem [{\citenamefont {Huh}\ \emph {et~al.}(2015)\citenamefont {Huh},
  \citenamefont {Guerreschi}, \citenamefont {Peropadre}, \citenamefont
  {McClean},\ and\ \citenamefont {Aspuru-Guzik}}]{Huh:2015}%
  \BibitemOpen
  \bibfield  {author} {\bibinfo {author} {\bibfnamefont {J.}~\bibnamefont
  {Huh}}, \bibinfo {author} {\bibfnamefont {G.~G.}\ \bibnamefont {Guerreschi}},
  \bibinfo {author} {\bibfnamefont {B.}~\bibnamefont {Peropadre}}, \bibinfo
  {author} {\bibfnamefont {J.~R.}\ \bibnamefont {McClean}}, \ and\ \bibinfo
  {author} {\bibfnamefont {A.}~\bibnamefont {Aspuru-Guzik}},\ }\href
  {http://dx.doi.org/10.1038/nphoton.2015.153} {\bibfield  {journal} {\bibinfo
  {journal} {Nat. Photon.}\ }\textbf {\bibinfo {volume} {9}},\ \bibinfo {pages}
  {615} (\bibinfo {year} {2015})}\BibitemShut {NoStop}%
\bibitem [{\citenamefont {Wecker}\ \emph {et~al.}(2015)\citenamefont {Wecker},
  \citenamefont {Hastings},\ and\ \citenamefont {Troyer}}]{Wecker:2015a}%
  \BibitemOpen
  \bibfield  {author} {\bibinfo {author} {\bibfnamefont {D.}~\bibnamefont
  {Wecker}}, \bibinfo {author} {\bibfnamefont {M.~B.}\ \bibnamefont
  {Hastings}}, \ and\ \bibinfo {author} {\bibfnamefont {M.}~\bibnamefont
  {Troyer}},\ }\href {\doibase 10.1103/PhysRevA.92.042303} {\bibfield
  {journal} {\bibinfo  {journal} {Phys. Rev. A}\ }\textbf {\bibinfo {volume}
  {92}},\ \bibinfo {pages} {042303} (\bibinfo {year} {2015})}\BibitemShut
  {NoStop}%
\bibitem [{\citenamefont {Mueck}(2015)}]{Mueck:2015}%
  \BibitemOpen
  \bibfield  {author} {\bibinfo {author} {\bibfnamefont {L.}~\bibnamefont
  {Mueck}},\ }\href {\doibase 10.1038/nchem.2248} {\bibfield  {journal}
  {\bibinfo  {journal} {Nat. Chem.}\ }\textbf {\bibinfo {volume} {7}},\
  \bibinfo {pages} {361} (\bibinfo {year} {2015})}\BibitemShut {NoStop}%
\bibitem [{\citenamefont {Abrams}\ and\ \citenamefont
  {Lloyd}(1997)}]{Abrams1997}%
  \BibitemOpen
  \bibfield  {author} {\bibinfo {author} {\bibfnamefont {D.~S.}\ \bibnamefont
  {Abrams}}\ and\ \bibinfo {author} {\bibfnamefont {S.}~\bibnamefont {Lloyd}},\
  }\href@noop {} {\bibfield  {journal} {\bibinfo  {journal} {Phys. Rev. Lett.}\
  }\textbf {\bibinfo {volume} {79}},\ \bibinfo {pages} {4} (\bibinfo {year}
  {1997})}\BibitemShut {NoStop}%
\bibitem [{\citenamefont {Abrams}\ and\ \citenamefont
  {Lloyd}(1999)}]{Abrams1999}%
  \BibitemOpen
  \bibfield  {author} {\bibinfo {author} {\bibfnamefont {D.~S.}\ \bibnamefont
  {Abrams}}\ and\ \bibinfo {author} {\bibfnamefont {S.}~\bibnamefont {Lloyd}},\
  }\href@noop {} {\bibfield  {journal} {\bibinfo  {journal} {Phys. Rev. Lett.}\
  }\textbf {\bibinfo {volume} {83}},\ \bibinfo {pages} {5162} (\bibinfo {year}
  {1999})}\BibitemShut {NoStop}%
\bibitem [{\citenamefont {Kitaev}(1997)}]{Kitaev:1997}%
  \BibitemOpen
  \bibfield  {author} {\bibinfo {author} {\bibfnamefont {A.~Y.}\ \bibnamefont
  {Kitaev}},\ }\href {http://dx.doi.org/10.1070/rm1997v052n06abeh002155}
  {\bibfield  {journal} {\bibinfo  {journal} {Russian Math. Surveys}\ }\textbf
  {\bibinfo {volume} {52}},\ \bibinfo {pages} {1191} (\bibinfo {year}
  {1997})}\BibitemShut {NoStop}%
\bibitem [{\citenamefont {McClean}\ \emph {et~al.}(2014)\citenamefont
  {McClean}, \citenamefont {Babbush}, \citenamefont {Love},\ and\ \citenamefont
  {Aspuru-Guzik}}]{McClean:2014}%
  \BibitemOpen
  \bibfield  {author} {\bibinfo {author} {\bibfnamefont {J.~R.}\ \bibnamefont
  {McClean}}, \bibinfo {author} {\bibfnamefont {R.}~\bibnamefont {Babbush}},
  \bibinfo {author} {\bibfnamefont {P.~J.}\ \bibnamefont {Love}}, \ and\
  \bibinfo {author} {\bibfnamefont {A.}~\bibnamefont {Aspuru-Guzik}},\
  }\href@noop {} {\bibfield  {journal} {\bibinfo  {journal} {J. Phys. Chem.
  Lett.}\ }\textbf {\bibinfo {volume} {5}},\ \bibinfo {pages} {4368} (\bibinfo
  {year} {2014})}\BibitemShut {NoStop}%
\bibitem [{\citenamefont {Babbush}\ \emph {et~al.}(2015)\citenamefont
  {Babbush}, \citenamefont {McClean}, \citenamefont {Wecker}, \citenamefont
  {Aspuru-Guzik},\ and\ \citenamefont {Wiebe}}]{Babbush:2015}%
  \BibitemOpen
  \bibfield  {author} {\bibinfo {author} {\bibfnamefont {R.}~\bibnamefont
  {Babbush}}, \bibinfo {author} {\bibfnamefont {J.~R.}\ \bibnamefont
  {McClean}}, \bibinfo {author} {\bibfnamefont {D.}~\bibnamefont {Wecker}},
  \bibinfo {author} {\bibfnamefont {A.}~\bibnamefont {Aspuru-Guzik}}, \ and\
  \bibinfo {author} {\bibfnamefont {N.}~\bibnamefont {Wiebe}},\ }\href
  {http://link.aps.org/doi/10.1103/PhysRevA.91.022311} {\bibfield  {journal}
  {\bibinfo  {journal} {Phys. Rev. A}\ }\textbf {\bibinfo {volume} {91}},\
  \bibinfo {pages} {022311} (\bibinfo {year} {2015})}\BibitemShut {NoStop}%
\bibitem [{\citenamefont {Hastings}\ \emph {et~al.}(2015)\citenamefont
  {Hastings}, \citenamefont {Wecker}, \citenamefont {Bauer},\ and\
  \citenamefont {Troyer}}]{Hastings2014}%
  \BibitemOpen
  \bibfield  {author} {\bibinfo {author} {\bibfnamefont {M.~B.}\ \bibnamefont
  {Hastings}}, \bibinfo {author} {\bibfnamefont {D.}~\bibnamefont {Wecker}},
  \bibinfo {author} {\bibfnamefont {B.}~\bibnamefont {Bauer}}, \ and\ \bibinfo
  {author} {\bibfnamefont {M.}~\bibnamefont {Troyer}},\ }\href@noop {}
  {\bibfield  {journal} {\bibinfo  {journal} {Quantum Inf. Comput.}\ }\textbf
  {\bibinfo {volume} {15}},\ \bibinfo {pages} {1} (\bibinfo {year}
  {2015})}\BibitemShut {NoStop}%
\bibitem [{\citenamefont {Poulin}\ \emph {et~al.}(2014)\citenamefont {Poulin},
  \citenamefont {Hastings}, \citenamefont {Wecker}, \citenamefont {Wiebe},
  \citenamefont {Doherty},\ and\ \citenamefont {Troyer}}]{Poulin2014}%
  \BibitemOpen
  \bibfield  {author} {\bibinfo {author} {\bibfnamefont {D.}~\bibnamefont
  {Poulin}}, \bibinfo {author} {\bibfnamefont {M.~B.}\ \bibnamefont
  {Hastings}}, \bibinfo {author} {\bibfnamefont {D.}~\bibnamefont {Wecker}},
  \bibinfo {author} {\bibfnamefont {N.}~\bibnamefont {Wiebe}}, \bibinfo
  {author} {\bibfnamefont {A.~C.}\ \bibnamefont {Doherty}}, \ and\ \bibinfo
  {author} {\bibfnamefont {M.}~\bibnamefont {Troyer}},\ }\href
  {http://arxiv.org/abs/1406.4920} {\bibfield  {journal} {\bibinfo  {journal}
  {ArXiv e-prints}\ } (\bibinfo {year} {2014})},\ \Eprint
  {http://arxiv.org/abs/1406.4920} {arXiv:1406.4920 [quant-ph]} \BibitemShut
  {NoStop}%
\bibitem [{\citenamefont {{Babbush}}\ \emph {et~al.}(2015)\citenamefont
  {{Babbush}}, \citenamefont {{Berry}}, \citenamefont {{Kivlichan}},
  \citenamefont {{Wei}}, \citenamefont {{Love}},\ and\ \citenamefont
  {{Aspuru-Guzik}}}]{Babbush:2015a}%
  \BibitemOpen
  \bibfield  {author} {\bibinfo {author} {\bibfnamefont {R.}~\bibnamefont
  {{Babbush}}}, \bibinfo {author} {\bibfnamefont {D.~W.}\ \bibnamefont
  {{Berry}}}, \bibinfo {author} {\bibfnamefont {I.~D.}\ \bibnamefont
  {{Kivlichan}}}, \bibinfo {author} {\bibfnamefont {A.~Y.}\ \bibnamefont
  {{Wei}}}, \bibinfo {author} {\bibfnamefont {P.~J.}\ \bibnamefont {{Love}}}, \
  and\ \bibinfo {author} {\bibfnamefont {A.}~\bibnamefont {{Aspuru-Guzik}}},\
  }\href@noop {} {\bibfield  {journal} {\bibinfo  {journal} {ArXiv e-prints}\ }
  (\bibinfo {year} {2015})},\ \Eprint {http://arxiv.org/abs/1506.01020}
  {arXiv:1506.01020 [quant-ph]} \BibitemShut {NoStop}%
\bibitem [{\citenamefont {Lanyon}\ \emph {et~al.}(2010)\citenamefont {Lanyon},
  \citenamefont {Whitfield}, \citenamefont {Gillett}, \citenamefont {Goggin},
  \citenamefont {Almeida}, \citenamefont {Kassal}, \citenamefont {Biamonte},
  \citenamefont {Mohseni}, \citenamefont {Powell}, \citenamefont {Barbieri},
  \citenamefont {Aspuru-Guzik},\ and\ \citenamefont {White}}]{Lanyon:2010}%
  \BibitemOpen
  \bibfield  {author} {\bibinfo {author} {\bibfnamefont {B.~P.}\ \bibnamefont
  {Lanyon}}, \bibinfo {author} {\bibfnamefont {J.~D.}\ \bibnamefont
  {Whitfield}}, \bibinfo {author} {\bibfnamefont {G.~G.}\ \bibnamefont
  {Gillett}}, \bibinfo {author} {\bibfnamefont {M.~E.}\ \bibnamefont {Goggin}},
  \bibinfo {author} {\bibfnamefont {M.~P.}\ \bibnamefont {Almeida}}, \bibinfo
  {author} {\bibfnamefont {I.}~\bibnamefont {Kassal}}, \bibinfo {author}
  {\bibfnamefont {J.~D.}\ \bibnamefont {Biamonte}}, \bibinfo {author}
  {\bibfnamefont {M.}~\bibnamefont {Mohseni}}, \bibinfo {author} {\bibfnamefont
  {B.~J.}\ \bibnamefont {Powell}}, \bibinfo {author} {\bibfnamefont
  {M.}~\bibnamefont {Barbieri}}, \bibinfo {author} {\bibfnamefont
  {A.}~\bibnamefont {Aspuru-Guzik}}, \ and\ \bibinfo {author} {\bibfnamefont
  {A.~G.}\ \bibnamefont {White}},\ }\href {http://dx.doi.org/10.1038/nchem.483}
  {\bibfield  {journal} {\bibinfo  {journal} {Nat. Chem.}\ }\textbf {\bibinfo
  {volume} {2}},\ \bibinfo {pages} {106} (\bibinfo {year} {2010})}\BibitemShut
  {NoStop}%
\bibitem [{\citenamefont {Lu}\ \emph {et~al.}(2011)\citenamefont {Lu},
  \citenamefont {Xu}, \citenamefont {Xu}, \citenamefont {Chen}, \citenamefont
  {Gong}, \citenamefont {Peng},\ and\ \citenamefont {Du}}]{Lu:2011}%
  \BibitemOpen
  \bibfield  {author} {\bibinfo {author} {\bibfnamefont {D.}~\bibnamefont
  {Lu}}, \bibinfo {author} {\bibfnamefont {N.}~\bibnamefont {Xu}}, \bibinfo
  {author} {\bibfnamefont {R.}~\bibnamefont {Xu}}, \bibinfo {author}
  {\bibfnamefont {H.}~\bibnamefont {Chen}}, \bibinfo {author} {\bibfnamefont
  {J.}~\bibnamefont {Gong}}, \bibinfo {author} {\bibfnamefont {X.}~\bibnamefont
  {Peng}}, \ and\ \bibinfo {author} {\bibfnamefont {J.}~\bibnamefont {Du}},\
  }\href {http://link.aps.org/doi/10.1103/PhysRevLett.107.020501} {\bibfield
  {journal} {\bibinfo  {journal} {Phys. Rev. Lett.}\ }\textbf {\bibinfo
  {volume} {107}},\ \bibinfo {pages} {020501} (\bibinfo {year}
  {2011})}\BibitemShut {NoStop}%
\bibitem [{\citenamefont {Aspuru-Guzik}\ and\ \citenamefont
  {Walther}(2012)}]{Walther:2012}%
  \BibitemOpen
  \bibfield  {author} {\bibinfo {author} {\bibfnamefont {A.}~\bibnamefont
  {Aspuru-Guzik}}\ and\ \bibinfo {author} {\bibfnamefont {P.}~\bibnamefont
  {Walther}},\ }\href {http://dx.doi.org/10.1038/nphys2253} {\bibfield
  {journal} {\bibinfo  {journal} {Nat. Phys.}\ }\textbf {\bibinfo {volume}
  {8}},\ \bibinfo {pages} {285} (\bibinfo {year} {2012})}\BibitemShut {NoStop}%
\bibitem [{\citenamefont {{Peruzzo}}\ \emph {et~al.}(2014)\citenamefont
  {{Peruzzo}}, \citenamefont {{McClean}}, \citenamefont {{Shadbolt}},
  \citenamefont {{Yung}}, \citenamefont {{Zhou}}, \citenamefont {{Love}},
  \citenamefont {{Aspuru-Guzik}},\ and\ \citenamefont
  {{O'Brien}}}]{Peruzzo2014}%
  \BibitemOpen
  \bibfield  {author} {\bibinfo {author} {\bibfnamefont {A.}~\bibnamefont
  {{Peruzzo}}}, \bibinfo {author} {\bibfnamefont {J.}~\bibnamefont
  {{McClean}}}, \bibinfo {author} {\bibfnamefont {P.}~\bibnamefont
  {{Shadbolt}}}, \bibinfo {author} {\bibfnamefont {M.-H.}\ \bibnamefont
  {{Yung}}}, \bibinfo {author} {\bibfnamefont {X.-Q.}\ \bibnamefont {{Zhou}}},
  \bibinfo {author} {\bibfnamefont {P.~J.}\ \bibnamefont {{Love}}}, \bibinfo
  {author} {\bibfnamefont {A.}~\bibnamefont {{Aspuru-Guzik}}}, \ and\ \bibinfo
  {author} {\bibfnamefont {J.~L.}\ \bibnamefont {{O'Brien}}},\ }\href
  {http://dx.doi.org/10.1038/ncomms5213} {\bibfield  {journal} {\bibinfo
  {journal} {Nat. Commun.}\ }\textbf {\bibinfo {volume} {5}} (\bibinfo {year}
  {2014})}\BibitemShut {NoStop}%
\bibitem [{\citenamefont {Wang}\ \emph {et~al.}(2015)\citenamefont {Wang},
  \citenamefont {Dolde}, \citenamefont {Biamonte}, \citenamefont {Babbush},
  \citenamefont {Bergholm}, \citenamefont {Yang}, \citenamefont {Jakobi},
  \citenamefont {Neumann}, \citenamefont {Aspuru-Guzik}, \citenamefont
  {Whitfield},\ and\ \citenamefont {Wrachtrup}}]{Wang:2015}%
  \BibitemOpen
  \bibfield  {author} {\bibinfo {author} {\bibfnamefont {Y.}~\bibnamefont
  {Wang}}, \bibinfo {author} {\bibfnamefont {F.}~\bibnamefont {Dolde}},
  \bibinfo {author} {\bibfnamefont {J.}~\bibnamefont {Biamonte}}, \bibinfo
  {author} {\bibfnamefont {R.}~\bibnamefont {Babbush}}, \bibinfo {author}
  {\bibfnamefont {V.}~\bibnamefont {Bergholm}}, \bibinfo {author}
  {\bibfnamefont {S.}~\bibnamefont {Yang}}, \bibinfo {author} {\bibfnamefont
  {I.}~\bibnamefont {Jakobi}}, \bibinfo {author} {\bibfnamefont
  {P.}~\bibnamefont {Neumann}}, \bibinfo {author} {\bibfnamefont
  {A.}~\bibnamefont {Aspuru-Guzik}}, \bibinfo {author} {\bibfnamefont {J.~D.}\
  \bibnamefont {Whitfield}}, \ and\ \bibinfo {author} {\bibfnamefont
  {J.}~\bibnamefont {Wrachtrup}},\ }\href@noop {} {\bibfield  {journal}
  {\bibinfo  {journal} {ACS Nano}\ } (\bibinfo {year} {2015})}\BibitemShut
  {NoStop}%
\bibitem [{\citenamefont {{Shen}}\ \emph {et~al.}(2015)\citenamefont {{Shen}},
  \citenamefont {{Zhang}}, \citenamefont {{Zhang}}, \citenamefont {{Zhang}},
  \citenamefont {{Yung}},\ and\ \citenamefont {{Kim}}}]{Shen:2015}%
  \BibitemOpen
  \bibfield  {author} {\bibinfo {author} {\bibfnamefont {Y.}~\bibnamefont
  {{Shen}}}, \bibinfo {author} {\bibfnamefont {X.}~\bibnamefont {{Zhang}}},
  \bibinfo {author} {\bibfnamefont {S.}~\bibnamefont {{Zhang}}}, \bibinfo
  {author} {\bibfnamefont {J.-N.}\ \bibnamefont {{Zhang}}}, \bibinfo {author}
  {\bibfnamefont {M.-H.}\ \bibnamefont {{Yung}}}, \ and\ \bibinfo {author}
  {\bibfnamefont {K.}~\bibnamefont {{Kim}}},\ }\href@noop {} {\bibfield
  {journal} {\bibinfo  {journal} {ArXiv e-prints}\ } (\bibinfo {year}
  {2015})},\ \Eprint {http://arxiv.org/abs/1506.00443} {arXiv:1506.00443
  [quant-ph]} \BibitemShut {NoStop}%
\bibitem [{\citenamefont {Barrett}\ \emph {et~al.}(2013)\citenamefont
  {Barrett}, \citenamefont {Hammerer}, \citenamefont {Harrison}, \citenamefont
  {Northup},\ and\ \citenamefont {Osborne}}]{Barrett:2013}%
  \BibitemOpen
  \bibfield  {author} {\bibinfo {author} {\bibfnamefont {S.}~\bibnamefont
  {Barrett}}, \bibinfo {author} {\bibfnamefont {K.}~\bibnamefont {Hammerer}},
  \bibinfo {author} {\bibfnamefont {S.}~\bibnamefont {Harrison}}, \bibinfo
  {author} {\bibfnamefont {T.~E.}\ \bibnamefont {Northup}}, \ and\ \bibinfo
  {author} {\bibfnamefont {T.~J.}\ \bibnamefont {Osborne}},\ }\href {\doibase
  10.1103/PhysRevLett.110.090501} {\bibfield  {journal} {\bibinfo  {journal}
  {Phys. Rev. Lett.}\ }\textbf {\bibinfo {volume} {110}},\ \bibinfo {pages}
  {090501} (\bibinfo {year} {2013})}\BibitemShut {NoStop}%
\bibitem [{\citenamefont {{McClean}}\ \emph {et~al.}(2015)\citenamefont
  {{McClean}}, \citenamefont {{Romero}}, \citenamefont {{Babbush}},\ and\
  \citenamefont {{Aspuru-Guzik}}}]{McClean:2015}%
  \BibitemOpen
  \bibfield  {author} {\bibinfo {author} {\bibfnamefont {J.~R.}\ \bibnamefont
  {{McClean}}}, \bibinfo {author} {\bibfnamefont {J.}~\bibnamefont {{Romero}}},
  \bibinfo {author} {\bibfnamefont {R.}~\bibnamefont {{Babbush}}}, \ and\
  \bibinfo {author} {\bibfnamefont {A.}~\bibnamefont {{Aspuru-Guzik}}},\
  }\href@noop {} {\bibfield  {journal} {\bibinfo  {journal} {ArXiv e-prints}\ }
  (\bibinfo {year} {2015})},\ \Eprint {http://arxiv.org/abs/1509.04279}
  {arXiv:1509.04279 [quant-ph]} \BibitemShut {NoStop}%
\bibitem [{\citenamefont {Yung}\ \emph {et~al.}(2014)\citenamefont {Yung},
  \citenamefont {Casanova}, \citenamefont {Mezzacapo}, \citenamefont {McClean},
  \citenamefont {Lamata}, \citenamefont {Aspuru-Guzik},\ and\ \citenamefont
  {Solano}}]{Yung:2014}%
  \BibitemOpen
  \bibfield  {author} {\bibinfo {author} {\bibfnamefont {M.~H.}\ \bibnamefont
  {Yung}}, \bibinfo {author} {\bibfnamefont {J.}~\bibnamefont {Casanova}},
  \bibinfo {author} {\bibfnamefont {A.}~\bibnamefont {Mezzacapo}}, \bibinfo
  {author} {\bibfnamefont {J.~R.}\ \bibnamefont {McClean}}, \bibinfo {author}
  {\bibfnamefont {L.}~\bibnamefont {Lamata}}, \bibinfo {author} {\bibfnamefont
  {A.}~\bibnamefont {Aspuru-Guzik}}, \ and\ \bibinfo {author} {\bibfnamefont
  {E.}~\bibnamefont {Solano}},\ }\href@noop {} {\bibfield  {journal} {\bibinfo
  {journal} {Sci. Rep.}\ }\textbf {\bibinfo {volume} {4}},\ \bibinfo {pages}
  {1} (\bibinfo {year} {2014})}\BibitemShut {NoStop}%
\bibitem [{\citenamefont {{Dallaire-Demers}}\ and\ \citenamefont
  {{Wilhelm}}(2015)}]{Dallaire:2015}%
  \BibitemOpen
  \bibfield  {author} {\bibinfo {author} {\bibfnamefont {P.-L.}\ \bibnamefont
  {{Dallaire-Demers}}}\ and\ \bibinfo {author} {\bibfnamefont {F.~K.}\
  \bibnamefont {{Wilhelm}}},\ }\href@noop {} {\bibfield  {journal} {\bibinfo
  {journal} {ArXiv e-prints}\ } (\bibinfo {year} {2015})},\ \Eprint
  {http://arxiv.org/abs/1508.04328} {arXiv:1508.04328 [quant-ph]} \BibitemShut
  {NoStop}%
\bibitem [{\citenamefont {{Bauer}}\ \emph {et~al.}(2015)\citenamefont
  {{Bauer}}, \citenamefont {{Wecker}}, \citenamefont {{Millis}}, \citenamefont
  {{Hastings}},\ and\ \citenamefont {{Troyer}}}]{Bauer:2015}%
  \BibitemOpen
  \bibfield  {author} {\bibinfo {author} {\bibfnamefont {B.}~\bibnamefont
  {{Bauer}}}, \bibinfo {author} {\bibfnamefont {D.}~\bibnamefont {{Wecker}}},
  \bibinfo {author} {\bibfnamefont {A.~J.}\ \bibnamefont {{Millis}}}, \bibinfo
  {author} {\bibfnamefont {M.~B.}\ \bibnamefont {{Hastings}}}, \ and\ \bibinfo
  {author} {\bibfnamefont {M.}~\bibnamefont {{Troyer}}},\ }\href@noop {}
  {\bibfield  {journal} {\bibinfo  {journal} {ArXiv e-prints}\ } (\bibinfo
  {year} {2015})},\ \Eprint {http://arxiv.org/abs/1510.03859} {arXiv:1510.03859
  [quant-ph]} \BibitemShut {NoStop}%
\bibitem [{\citenamefont {{O'Malley}}\ \emph {et~al.}(2015)\citenamefont
  {{O'Malley}}, \citenamefont {{Babbush}}, \citenamefont {{Kivlichan}},
  \citenamefont {{Romero}}, \citenamefont {{McClean}}, \citenamefont
  {{Barends}}, \citenamefont {{Kelly}}, \citenamefont {{Roushan}},
  \citenamefont {{Tranter}}, \citenamefont {{Ding}}, \citenamefont
  {{Campbell}}, \citenamefont {{Chen}}, \citenamefont {{Chen}}, \citenamefont
  {{Chiaro}}, \citenamefont {{Dunsworth}}, \citenamefont {{Fowler}},
  \citenamefont {{Jeffrey}}, \citenamefont {{Megrant}}, \citenamefont
  {{Mutus}}, \citenamefont {{Neill}}, \citenamefont {{Quintana}}, \citenamefont
  {{Sank}}, \citenamefont {{Vainsencher}}, \citenamefont {{Wenner}},
  \citenamefont {{White}}, \citenamefont {{Coveney}}, \citenamefont {{Love}},
  \citenamefont {{Neven}}, \citenamefont {{Aspuru-Guzik}},\ and\ \citenamefont
  {{Martinis}}}]{OMalley:2015}%
  \BibitemOpen
  \bibfield  {author} {\bibinfo {author} {\bibfnamefont {P.~J.~J.}\
  \bibnamefont {{O'Malley}}}, \bibinfo {author} {\bibfnamefont
  {R.}~\bibnamefont {{Babbush}}}, \bibinfo {author} {\bibfnamefont {I.~D.}\
  \bibnamefont {{Kivlichan}}}, \bibinfo {author} {\bibfnamefont
  {J.}~\bibnamefont {{Romero}}}, \bibinfo {author} {\bibfnamefont {J.~R.}\
  \bibnamefont {{McClean}}}, \bibinfo {author} {\bibfnamefont {R.}~\bibnamefont
  {{Barends}}}, \bibinfo {author} {\bibfnamefont {J.}~\bibnamefont {{Kelly}}},
  \bibinfo {author} {\bibfnamefont {P.}~\bibnamefont {{Roushan}}}, \bibinfo
  {author} {\bibfnamefont {A.}~\bibnamefont {{Tranter}}}, \bibinfo {author}
  {\bibfnamefont {N.}~\bibnamefont {{Ding}}}, \bibinfo {author} {\bibfnamefont
  {B.}~\bibnamefont {{Campbell}}}, \bibinfo {author} {\bibfnamefont
  {Y.}~\bibnamefont {{Chen}}}, \bibinfo {author} {\bibfnamefont
  {Z.}~\bibnamefont {{Chen}}}, \bibinfo {author} {\bibfnamefont
  {B.}~\bibnamefont {{Chiaro}}}, \bibinfo {author} {\bibfnamefont
  {A.}~\bibnamefont {{Dunsworth}}}, \bibinfo {author} {\bibfnamefont {A.~G.}\
  \bibnamefont {{Fowler}}}, \bibinfo {author} {\bibfnamefont {E.}~\bibnamefont
  {{Jeffrey}}}, \bibinfo {author} {\bibfnamefont {A.}~\bibnamefont
  {{Megrant}}}, \bibinfo {author} {\bibfnamefont {J.~Y.}\ \bibnamefont
  {{Mutus}}}, \bibinfo {author} {\bibfnamefont {C.}~\bibnamefont {{Neill}}},
  \bibinfo {author} {\bibfnamefont {C.}~\bibnamefont {{Quintana}}}, \bibinfo
  {author} {\bibfnamefont {D.}~\bibnamefont {{Sank}}}, \bibinfo {author}
  {\bibfnamefont {A.}~\bibnamefont {{Vainsencher}}}, \bibinfo {author}
  {\bibfnamefont {J.}~\bibnamefont {{Wenner}}}, \bibinfo {author}
  {\bibfnamefont {T.~C.}\ \bibnamefont {{White}}}, \bibinfo {author}
  {\bibfnamefont {P.~V.}\ \bibnamefont {{Coveney}}}, \bibinfo {author}
  {\bibfnamefont {P.~J.}\ \bibnamefont {{Love}}}, \bibinfo {author}
  {\bibfnamefont {H.}~\bibnamefont {{Neven}}}, \bibinfo {author} {\bibfnamefont
  {A.}~\bibnamefont {{Aspuru-Guzik}}}, \ and\ \bibinfo {author} {\bibfnamefont
  {J.~M.}\ \bibnamefont {{Martinis}}},\ }\href@noop {} {\bibfield  {journal}
  {\bibinfo  {journal} {ArXiv e-prints}\ } (\bibinfo {year} {2015})},\ \Eprint
  {http://arxiv.org/abs/1512.06860} {arXiv:1512.06860 [quant-ph]} \BibitemShut
  {NoStop}%
\bibitem [{\citenamefont {Kraus}\ \emph {et~al.}(1983)\citenamefont {Kraus},
  \citenamefont {Böhm}, \citenamefont {Dollard},\ and\ \citenamefont
  {Wootters}}]{Kraus:1983}%
  \BibitemOpen
  \bibinfo {editor} {\bibfnamefont {K.}~\bibnamefont {Kraus}}, \bibinfo
  {editor} {\bibfnamefont {A.}~\bibnamefont {Böhm}}, \bibinfo {editor}
  {\bibfnamefont {J.~D.}\ \bibnamefont {Dollard}}, \ and\ \bibinfo {editor}
  {\bibfnamefont {W.~H.}\ \bibnamefont {Wootters}},\ eds.,\ \href {\doibase
  10.1007/3-540-12732-1} {\emph {\bibinfo {title} {States, Effects, and
  Operations Fundamental Notions of Quantum Theory}}}\ (\bibinfo  {publisher}
  {Springer Berlin Heidelberg},\ \bibinfo {year} {1983})\BibitemShut {NoStop}%
\bibitem [{\citenamefont {Watermann}\ \emph {et~al.}(2014)\citenamefont
  {Watermann}, \citenamefont {Scherrer},\ and\ \citenamefont
  {Sebastiani}}]{Volker:2014}%
  \BibitemOpen
  \bibfield  {author} {\bibinfo {author} {\bibfnamefont {T.}~\bibnamefont
  {Watermann}}, \bibinfo {author} {\bibfnamefont {A.}~\bibnamefont {Scherrer}},
  \ and\ \bibinfo {author} {\bibfnamefont {D.}~\bibnamefont {Sebastiani}},\
  }in\ \href {\doibase 10.1007/978-3-319-06379-9_5} {\emph {\bibinfo
  {booktitle} {Many-Electron Approaches in Physics, Chemistry and
  Mathematics}}},\ \bibinfo {series and number} {Mathematical Physics
  Studies},\ \bibinfo {editor} {edited by\ \bibinfo {editor} {\bibfnamefont
  {V.}~\bibnamefont {Bach}}\ and\ \bibinfo {editor} {\bibfnamefont
  {L.}~\bibnamefont {Delle~Site}}}\ (\bibinfo  {publisher} {Springer
  International Publishing},\ \bibinfo {year} {2014})\ pp.\ \bibinfo {pages}
  {97--110}\BibitemShut {NoStop}%
\bibitem [{\citenamefont {Hehre}(1969)}]{Hehre:1969}%
  \BibitemOpen
  \bibfield  {author} {\bibinfo {author} {\bibfnamefont {W.~J.}\ \bibnamefont
  {Hehre}},\ }\href {\doibase 10.1063/1.1672392} {\bibfield  {journal}
  {\bibinfo  {journal} {J. Chem. Phys.}\ }\textbf {\bibinfo {volume} {51}},\
  \bibinfo {pages} {2657} (\bibinfo {year} {1969})}\BibitemShut {NoStop}%
\bibitem [{\citenamefont {Jordan}\ and\ \citenamefont
  {Wigner}(1928)}]{Jordan1928}%
  \BibitemOpen
  \bibfield  {author} {\bibinfo {author} {\bibfnamefont {P.}~\bibnamefont
  {Jordan}}\ and\ \bibinfo {author} {\bibfnamefont {E.}~\bibnamefont
  {Wigner}},\ }\href@noop {} {\bibfield  {journal} {\bibinfo  {journal}
  {Zeitschrift f\"{u}r Physik}\ }\textbf {\bibinfo {volume} {47}},\ \bibinfo
  {pages} {631} (\bibinfo {year} {1928})}\BibitemShut {NoStop}%
\bibitem [{\citenamefont {Kassal}\ \emph {et~al.}(2008)\citenamefont {Kassal},
  \citenamefont {Jordan}, \citenamefont {Love}, \citenamefont {Mohseni},\ and\
  \citenamefont {Aspuru-Guzik}}]{Kassal:2008}%
  \BibitemOpen
  \bibfield  {author} {\bibinfo {author} {\bibfnamefont {I.}~\bibnamefont
  {Kassal}}, \bibinfo {author} {\bibfnamefont {S.~P.}\ \bibnamefont {Jordan}},
  \bibinfo {author} {\bibfnamefont {P.~J.}\ \bibnamefont {Love}}, \bibinfo
  {author} {\bibfnamefont {M.}~\bibnamefont {Mohseni}}, \ and\ \bibinfo
  {author} {\bibfnamefont {A.}~\bibnamefont {Aspuru-Guzik}},\ }\href
  {http://dx.doi.org/10.1073/pnas.0808245105} {\bibfield  {journal} {\bibinfo
  {journal} {Proc. Natl. Acad. Sci. U.S.A.}\ }\textbf {\bibinfo {volume}
  {105}},\ \bibinfo {pages} {18681} (\bibinfo {year} {2008})}\BibitemShut
  {NoStop}%
\bibitem [{\citenamefont {Ward}\ \emph {et~al.}(2009)\citenamefont {Ward},
  \citenamefont {Kassal},\ and\ \citenamefont {Aspuru-Guzik}}]{Ward:2009}%
  \BibitemOpen
  \bibfield  {author} {\bibinfo {author} {\bibfnamefont {N.~J.}\ \bibnamefont
  {Ward}}, \bibinfo {author} {\bibfnamefont {I.}~\bibnamefont {Kassal}}, \ and\
  \bibinfo {author} {\bibfnamefont {A.}~\bibnamefont {Aspuru-Guzik}},\ }\href
  {http://dx.doi.org/10.1063/1.3115177} {\bibfield  {journal} {\bibinfo
  {journal} {J. Chem. Phys.}\ }\textbf {\bibinfo {volume} {130}},\ \bibinfo
  {pages} {194105} (\bibinfo {year} {2009})}\BibitemShut {NoStop}%
\bibitem [{\citenamefont {{Toloui}}\ and\ \citenamefont
  {{Love}}(2013)}]{Toloui:2013}%
  \BibitemOpen
  \bibfield  {author} {\bibinfo {author} {\bibfnamefont {B.}~\bibnamefont
  {{Toloui}}}\ and\ \bibinfo {author} {\bibfnamefont {P.~J.}\ \bibnamefont
  {{Love}}},\ }\href@noop {} {\bibfield  {journal} {\bibinfo  {journal} {ArXiv
  e-prints}\ } (\bibinfo {year} {2013})},\ \Eprint
  {http://arxiv.org/abs/1312.2579} {arXiv:1312.2579 [quant-ph]} \BibitemShut
  {NoStop}%
\bibitem [{\citenamefont {Welch}\ \emph {et~al.}(2014)\citenamefont {Welch},
  \citenamefont {Greenbaum}, \citenamefont {Mostame},\ and\ \citenamefont
  {Aspuru-Guzik}}]{Welch:2014}%
  \BibitemOpen
  \bibfield  {author} {\bibinfo {author} {\bibfnamefont {J.}~\bibnamefont
  {Welch}}, \bibinfo {author} {\bibfnamefont {D.}~\bibnamefont {Greenbaum}},
  \bibinfo {author} {\bibfnamefont {S.}~\bibnamefont {Mostame}}, \ and\
  \bibinfo {author} {\bibfnamefont {A.}~\bibnamefont {Aspuru-Guzik}},\ }\href
  {http://dx.doi.org/10.1088/1367-2630/16/3/033040} {\bibfield  {journal}
  {\bibinfo  {journal} {New J. Phys.}\ }\textbf {\bibinfo {volume} {16}},\
  \bibinfo {pages} {033040} (\bibinfo {year} {2014})}\BibitemShut {NoStop}%
\bibitem [{\citenamefont {{Whitfield}}(2015)}]{Whitfield:2015a}%
  \BibitemOpen
  \bibfield  {author} {\bibinfo {author} {\bibfnamefont {J.~D.}\ \bibnamefont
  {{Whitfield}}},\ }\href@noop {} {\bibfield  {journal} {\bibinfo  {journal}
  {ArXiv e-prints}\ } (\bibinfo {year} {2015})},\ \Eprint
  {http://arxiv.org/abs/1502.03771} {arXiv:1502.03771 [quant-ph]} \BibitemShut
  {NoStop}%
\bibitem [{\citenamefont {Helgaker}\ \emph {et~al.}(2014)\citenamefont
  {Helgaker}, \citenamefont {Jorgensen},\ and\ \citenamefont
  {Olsen}}]{Helgaker:2014}%
  \BibitemOpen
  \bibfield  {author} {\bibinfo {author} {\bibfnamefont {T.}~\bibnamefont
  {Helgaker}}, \bibinfo {author} {\bibfnamefont {P.}~\bibnamefont {Jorgensen}},
  \ and\ \bibinfo {author} {\bibfnamefont {J.}~\bibnamefont {Olsen}},\
  }\href@noop {} {\emph {\bibinfo {title} {Molecular electronic-structure
  theory}}}\ (\bibinfo  {publisher} {John Wiley \& Sons},\ \bibinfo {year}
  {2014})\BibitemShut {NoStop}%
\bibitem [{\citenamefont {Somma}\ \emph {et~al.}(2002)\citenamefont {Somma},
  \citenamefont {Ortiz}, \citenamefont {Gubernatis}, \citenamefont {Knill},\
  and\ \citenamefont {Laflamme}}]{Somma2002}%
  \BibitemOpen
  \bibfield  {author} {\bibinfo {author} {\bibfnamefont {R.}~\bibnamefont
  {Somma}}, \bibinfo {author} {\bibfnamefont {G.}~\bibnamefont {Ortiz}},
  \bibinfo {author} {\bibfnamefont {J.}~\bibnamefont {Gubernatis}}, \bibinfo
  {author} {\bibfnamefont {E.}~\bibnamefont {Knill}}, \ and\ \bibinfo {author}
  {\bibfnamefont {R.}~\bibnamefont {Laflamme}},\ }\href@noop {} {\bibfield
  {journal} {\bibinfo  {journal} {Phys. Rev. A}\ }\textbf {\bibinfo {volume}
  {65}},\ \bibinfo {pages} {17} (\bibinfo {year} {2002})}\BibitemShut {NoStop}%
\bibitem [{\citenamefont {Bravyi}\ and\ \citenamefont
  {Kitaev}(2000)}]{Bravyi2000}%
  \BibitemOpen
  \bibfield  {author} {\bibinfo {author} {\bibfnamefont {S.}~\bibnamefont
  {Bravyi}}\ and\ \bibinfo {author} {\bibfnamefont {A.}~\bibnamefont
  {Kitaev}},\ }\href {http://arxiv.org/abs/quant-ph/0003137} {\bibfield
  {journal} {\bibinfo  {journal} {Ann. Phys.}\ }\textbf {\bibinfo {volume}
  {298}},\ \bibinfo {pages} {18} (\bibinfo {year} {2000})}\BibitemShut
  {NoStop}%
\bibitem [{\citenamefont {Seeley}\ \emph {et~al.}(2012)\citenamefont {Seeley},
  \citenamefont {Richard},\ and\ \citenamefont {Love}}]{Seeley2012}%
  \BibitemOpen
  \bibfield  {author} {\bibinfo {author} {\bibfnamefont {J.~T.}\ \bibnamefont
  {Seeley}}, \bibinfo {author} {\bibfnamefont {M.~J.}\ \bibnamefont {Richard}},
  \ and\ \bibinfo {author} {\bibfnamefont {P.~J.}\ \bibnamefont {Love}},\
  }\href
  {http://scitation.aip.org/content/aip/journal/jcp/137/22/10.1063/1.4768229}
  {\bibfield  {journal} {\bibinfo  {journal} {J. Chem. Phys.}\ }\textbf
  {\bibinfo {volume} {137}},\ \bibinfo {eid} {224109} (\bibinfo {year}
  {2012})}\BibitemShut {NoStop}%
\bibitem [{\citenamefont {Tranter}\ \emph {et~al.}(2015)\citenamefont
  {Tranter}, \citenamefont {Sofia}, \citenamefont {Seeley}, \citenamefont
  {Kaicher}, \citenamefont {McClean}, \citenamefont {Babbush}, \citenamefont
  {Coveney}, \citenamefont {Mintert}, \citenamefont {Wilhelm},\ and\
  \citenamefont {Love}}]{Tranter:2015}%
  \BibitemOpen
  \bibfield  {author} {\bibinfo {author} {\bibfnamefont {A.}~\bibnamefont
  {Tranter}}, \bibinfo {author} {\bibfnamefont {S.}~\bibnamefont {Sofia}},
  \bibinfo {author} {\bibfnamefont {J.}~\bibnamefont {Seeley}}, \bibinfo
  {author} {\bibfnamefont {M.}~\bibnamefont {Kaicher}}, \bibinfo {author}
  {\bibfnamefont {J.}~\bibnamefont {McClean}}, \bibinfo {author} {\bibfnamefont
  {R.}~\bibnamefont {Babbush}}, \bibinfo {author} {\bibfnamefont {P.~V.}\
  \bibnamefont {Coveney}}, \bibinfo {author} {\bibfnamefont {F.}~\bibnamefont
  {Mintert}}, \bibinfo {author} {\bibfnamefont {F.}~\bibnamefont {Wilhelm}}, \
  and\ \bibinfo {author} {\bibfnamefont {P.~J.}\ \bibnamefont {Love}},\ }\href
  {\doibase 10.1002/qua.24969} {\bibfield  {journal} {\bibinfo  {journal} {Int.
  J. Quant. Chem.}\ }\textbf {\bibinfo {volume} {115}},\ \bibinfo {pages}
  {1431} (\bibinfo {year} {2015})}\BibitemShut {NoStop}%
\bibitem [{\citenamefont {{Moll}}\ \emph {et~al.}(2015)\citenamefont {{Moll}},
  \citenamefont {{Fuhrer}}, \citenamefont {{Staar}},\ and\ \citenamefont
  {{Tavernelli}}}]{Moll:2015}%
  \BibitemOpen
  \bibfield  {author} {\bibinfo {author} {\bibfnamefont {N.}~\bibnamefont
  {{Moll}}}, \bibinfo {author} {\bibfnamefont {A.}~\bibnamefont {{Fuhrer}}},
  \bibinfo {author} {\bibfnamefont {P.}~\bibnamefont {{Staar}}}, \ and\
  \bibinfo {author} {\bibfnamefont {I.}~\bibnamefont {{Tavernelli}}},\
  }\href@noop {} {\bibfield  {journal} {\bibinfo  {journal} {ArXiv e-prints}\ }
  (\bibinfo {year} {2015})},\ \Eprint {http://arxiv.org/abs/1510.04048}
  {arXiv:1510.04048 [quant-ph]} \BibitemShut {NoStop}%
\bibitem [{\citenamefont {Weinstein}(1934)}]{Weinstein:1934}%
  \BibitemOpen
  \bibfield  {author} {\bibinfo {author} {\bibfnamefont {D.}~\bibnamefont
  {Weinstein}},\ }\href@noop {} {\bibfield  {journal} {\bibinfo  {journal}
  {Proc. Natl. Acad. Sci. U.S.A.}\ }\textbf {\bibinfo {volume} {20}},\ \bibinfo
  {pages} {529} (\bibinfo {year} {1934})}\BibitemShut {NoStop}%
\bibitem [{\citenamefont {MacDonald}(1934)}]{MacDonald:1934}%
  \BibitemOpen
  \bibfield  {author} {\bibinfo {author} {\bibfnamefont {J.}~\bibnamefont
  {MacDonald}},\ }\href@noop {} {\bibfield  {journal} {\bibinfo  {journal}
  {Phys. Rev.}\ }\textbf {\bibinfo {volume} {46}},\ \bibinfo {pages} {828}
  (\bibinfo {year} {1934})}\BibitemShut {NoStop}%
\bibitem [{\citenamefont {{Bookatz}}\ \emph {et~al.}(2014)\citenamefont
  {{Bookatz}}, \citenamefont {{Farhi}},\ and\ \citenamefont
  {{Zhou}}}]{Bookatz:2014}%
  \BibitemOpen
  \bibfield  {author} {\bibinfo {author} {\bibfnamefont {A.~D.}\ \bibnamefont
  {{Bookatz}}}, \bibinfo {author} {\bibfnamefont {E.}~\bibnamefont {{Farhi}}},
  \ and\ \bibinfo {author} {\bibfnamefont {L.}~\bibnamefont {{Zhou}}},\
  }\href@noop {} {\bibfield  {journal} {\bibinfo  {journal} {ArXiv e-prints}\ }
  (\bibinfo {year} {2014})},\ \Eprint {http://arxiv.org/abs/1407.1485}
  {arXiv:1407.1485 [quant-ph]} \BibitemShut {NoStop}%
\bibitem [{\citenamefont {Mazziotti}(2003)}]{Mazziotti:2003}%
  \BibitemOpen
  \bibfield  {author} {\bibinfo {author} {\bibfnamefont {D.~A.}\ \bibnamefont
  {Mazziotti}},\ }\href@noop {} {\bibfield  {journal} {\bibinfo  {journal}
  {Physical Review A}\ }\textbf {\bibinfo {volume} {68}},\ \bibinfo {pages}
  {052501} (\bibinfo {year} {2003})}\BibitemShut {NoStop}%
\bibitem [{\citenamefont {Greenman}\ and\ \citenamefont
  {Mazziotti}(2008)}]{Greenman:2008}%
  \BibitemOpen
  \bibfield  {author} {\bibinfo {author} {\bibfnamefont {L.}~\bibnamefont
  {Greenman}}\ and\ \bibinfo {author} {\bibfnamefont {D.~A.}\ \bibnamefont
  {Mazziotti}},\ }\href@noop {} {\bibfield  {journal} {\bibinfo  {journal} {J.
  Chem. Phys.}\ }\textbf {\bibinfo {volume} {128}},\ \bibinfo {pages} {114109}
  (\bibinfo {year} {2008})}\BibitemShut {NoStop}%
\end{thebibliography}%

\end{document}